\documentclass{aa}  

\bibpunct{(}{)}{;}{a}{}{,}

\usepackage{graphicx, graphics}
\usepackage{txfonts}

\usepackage{color}

\begin{document} 

   \title{Chemistry of a newly detected circumbinary disk in Ophiuchus}

   \author{Elizabeth Artur de la Villarmois
          \inst{1} \and
          Lars E. Kristensen
          \inst{1} \and
          Jes K. J{\o}rgensen
          \inst{1} \and
         Edwin A. Bergin
          \inst{2} \and
         Christian Brinch          
         \inst{3} \and
         S{\o}ren Frimann
          \inst{4} \and
          Daniel Harsono
          \inst{5} \and
           Nami Sakai
          \inst{6} \and
          Satoshi Yamamoto
          \inst{7}
          }

\institute{Centre for Star and Planet Formation, Niels Bohr Institute $\&$ Natural History Museum of Denmark, University of Copenhagen, 
   {\O}ster Voldgade 5--7, 1350 Copenhagen K., Denmark\\
              \email{elizabeth.artur@nbi.ku.dk}
\and Department of Astronomy, University of Michigan, 311 West Hall, 1085 S. University Ave, Ann Arbor, MI 48109, USA
\and Niels Bohr International Academy, The Niels Bohr Institute, University of Copenhagen, Blegdamsvej 17, DK-2100 Copenhagen {\O}. Denmark
\and ICREA and Institut de Ci\`{e}ncies del Cosmos, Universitat de Barcelona, IEEC-UB, Mart\'{i} Franqu\`{e}s 1, 08028 Barcelona, Spain
\and Leiden Observatory, Leiden University, PO Box 9513, NL-2300 RA Leiden, the Netherlands
\and The Institute of Physical and Chemical Research (RIKEN), 2-1 Hirosawa, Wako-shi, Saitama 351-0198, Japan
\and Department of Physics, The University of Tokyo, Bunkyo-ku, Tokyo 113-0033, Japan
             }

  \abstract
    {Astronomers recently started discovering exoplanets around binary systems. Therefore, understanding the formation and evolution of circumbinary disks and their environment is crucial for a complete scenario of planet formation.}
    {The purpose of this paper is to present the detection of a circumbinary disk around the system Oph-IRS67 and analyse its chemical and physical structure.}
    {We present high-angular-resolution (0\farcs4, $\sim$60~AU) observations of C$^{17}$O, H$^{13}$CO$^{+}$, C$^{34}$S, SO$_{2}$, C$_{2}$H and c$-$C$_{3}$H$_{2}$ molecular transitions with the Atacama Large Millimeter/submillimeter Array (ALMA) at wavelengths of 0.8~mm. The spectrally and spatially resolved maps reveal the kinematics of the circumbinary disk as well as its chemistry. Molecular abundances are estimated using the non-local thermodynamic equilibrium (LTE) radiative-transfer tool RADEX.}
    {The continuum emission agrees with the position of Oph-IRS67 A and B, and reveals the presence of a circumbinary disk around the two sources. The circumbinary disk has a diameter of $\sim$620~AU and is well traced by C$^{17}$O and H$^{13}$CO$^{+}$ emission. Two further molecular species, C$_{2}$H and c$-$C$_{3}$H$_{2}$, trace a higher-density region which is spatially offset from the sources ($\sim$430~AU). Finally, SO$_{2}$ shows compact and broad emission around only one of the sources, Oph-IRS67 B. The molecular transitions which trace the circumbinary disk are consistent with a Keplerian profile on smaller disk scales ($\lesssim$~200 AU) and an infalling profile for larger envelope scales ($\gtrsim$~200 AU). The Keplerian fit leads to an enclosed mass of 2.2~M$_{\odot}$. Inferred CO abundances with respect to H$_{2}$ are comparable to the canonical ISM value of 2.7 $\times$ 10$^{-4}$, reflecting that freeze-out of CO in the disk midplane is not significant.}
     {Molecular emission and kinematic studies prove the existence and first detection of the circumbinary disk associated with the system Oph-IRS67. The high-density region shows a different chemistry than the disk, being enriched in carbon chain molecules. The lack of methanol emission agrees with the scenario where the extended disk dominates the mass budget in the innermost regions of the protostellar envelope, generating a flat density profile where less material is exposed to high temperatures, and thus, complex organic molecules would be associated with lower column densities. Finally, Oph-IRS67 is a promising candidate for proper motion studies and the detection of both circumstellar disks with higher-angular-resolution observations.}


   \maketitle

\section{Introduction}

Low-mass star formation takes place within dense cold molecular clouds, where individual cores collapse due to gravity. Because of the initial core rotation, conservation of angular momentum will lead to the formation of a circumstellar disk around the young stars. The angular momentum is transferred to the outer regions of the disk and moved outward by energetic outflows, sweeping away material from the envelope. In the earliest stages, the young source is deeply embedded in its infalling envelope of cold gas and dust, obscuring the radiation from the central star. As the system evolves, the envelope dissipates, revealing the pre-main sequence star and the circumstellar disk. The properties and evolution of these disks are crucial for the final mass of the host star and the initial conditions of planetary systems.

Within the envelope, the large variations in temperature (tens to hundreds of K) and density ($10^{5}$--$10^{9}$~cm$^{-3}$) leave strong chemical signatures that can potentially be observed directly, thus making these systems interesting laboratories for astrochemical studies \citep[e.g.][]{Jorgensen2004c}. Since individual molecular transitions are enhanced at specific temperatures and densities, their emission (and/or absorption) reveals the peculiarities and characteristics of the gas and can be used to reveal the physical structure and evolution of the young protostar (when the chemistry is taken into account). 

While the overall framework of low-mass star formation is well accepted nowadays, the details are significantly more complex, particularly when the formation of binary and/or multiple systems is considered \citep[e.g.][]{Tobin2016}. Circumstellar disks have been detected around individual binary components, as well as a circumbinary disk surrounding the system for Class I sources. Some cases show an alignment between the disks \citep[e.g.][]{Dutrey2015}, while other systems consist of misaligned components \citep[e.g.][]{Brinch2016, Takakuwa2017}. For more evolved systems, discrepancies are reported between disk sizes inferred from observations in multiple systems and predictions from tidal interaction models \citep[e.g.][]{Harris2012}. Apart from the physical structure, the chemical differentiation between the younger Class 0 and the more evolved Class I multiple systems provides information about the evolution of important parameters, such as temperature and density. Consequently, more observations are needed in order to understand the complexity of binary and multiple systems.

With the high angular resolution of the Atacama Large Millimeter/submillimeter array (ALMA), smaller scales ($\lesssim$10--50~AU) can be resolved towards nearby star-forming regions, providing very detailed observations of the environments in which stars form. One of the closest star-forming regions is the Ophiuchus molecular cloud \citep{Wilking2008}, characterised by a high visual extinction (\textit{A$_\mathrm{V}$}) of between 50 and 100 magnitudes, and  a high number of solar-type young stellar objects (YSOs) in different evolutionary stages \citep{Wilking2008}. Ophiuchus is therefore one of the most important regions for studies of the star formation process and early evolution of young stars.

One of the particularly interesting YSOs in Ophiuchus is the proto-binary system IRS67AB, located in the L1689 part of the cloud. This source is also known as L1689SAB and WLY2-67. The distance to IRS67AB and L1689 were recently determined through proper motion and trigonometric parallax measurements by \cite{OrtizLeon2017}. The distance was determined to be 147.3~$\pm$~3.4~pc to L1689 as a whole and 151.2~$\pm$~2.0~pc to IRS67AB specifically; in this paper, we adopt the latter. 

IRS67AB was associated with a large-scale outflow structure ($\gtrsim$~1000~AU), detected in CO \textit{J}=2$-$1 emission by \citet{Bontemps1996}, and was also surveyed at infrared and other wavelengths as part of the \textit{Spitzer} Space Telescope c2d legacy program \citep{Evans2009}. In that survey, the source was associated with a bolometric temperature (\textit{T$_\mathrm{bol}$}) of 130~K, bolometric luminosity (\textit{L$_\mathrm{bol}$}) of 4.0~L$_{\odot}$, infrared spectral index ($\alpha$$_\mathrm{IR}$) of 1.39 and visual extinction of 9.8 magnitudes, all characteristics of embedded Class I young stellar objects. Later, \citet{McClure2010} proved the binary nature of the system (L1689S-A and L1689S-B in their work), through infrared observations, calculated a separation of 0.6$''$ ($\sim$90~AU) between the two sources, and identified sources A and B as a disk and envelope candidate, respectively. 

In this paper, we present ALMA observations of Oph-IRS67AB. The study of this binary system demonstrates the complexity of protoplanetary disk formation and evolution (both circumstellar and circumbinary), from an observational point of view. Section 2 describes the observational procedure, data calibration, and covered molecular transitions. In Sect. 3, we present the results, highlighting the detection of the circumbinary disk and the chemical diversity of the system. Section 4 is dedicated to the analysis of the data, where dust and gas masses are calculated, and different velocity profiles are considered for the circumbinary disk. In addition, relative molecular abundances are estimated in order to compare them with values associated with a younger system. We discuss the structure and kinematics of the whole system, and the temperature profile of the circumbinary disk in Sect. 5. Finally, we end the paper with a summary in Sect. 6.

\section{Observations}
IRS67 was observed with ALMA on four occasions between 2015 May 21 and June 5 as part of a larger program to survey the line and continuum emission towards 12 Class~I protostars in Ophiuchus (program code: 2013.1.00955.S; PI: Jes J{\o}rgensen). At the time of the observations, 36 antennas were available in the array (37 for the June 5 observations) providing baselines between 21 and 556 metres (784 metres for the June 5 observations). Each of the four sessions provided an on-source time of 43 minutes in total for the 12 different sources (i.e. each source was observed for approximately 15~minutes in total)

The observations targeted five different spectral windows and the choice of species has been made specifically to trace different aspects of the structure of protostars. For example, the lines of C$^{17}$O, H$^{13}$CO$^{+}$ and C$^{34}$S are optically thin tracers of the kinematics of disk formation, while SO$_{2}$ and CH$_{3}$OH are expected to trace the warm chemistry in the inner envelope or disk. Two spectral windows consist of 960~channels, each with 122.07~kHz (0.11~km~s$^{-1}$) spectral resolution centred on C$^{17}$O \textit{J}=3$-$2 and C$^{34}$S \textit{J}=7$-$6, while the other three contain 1920~channels each with 244.14~kHz (0.22~km~s$^{-1}$) spectral resolution centred on H$^{13}$CO$^+$ \textit{J}=4$-$3, the CH$_3$OH \textit{J$_{k}$}=7$_k$$-$6$_k$ branch at 338.4~GHz and the CH$_3$CN 14$-$13 branch at 349.1~GHz. The latter two organic molecules are not detected towards IRS67, but instead the two settings pick up SO$_2$, C$_2$H, and c$-$C$_{3}$H$_{2}$ transitions. The spectral setup and covered molecular transitions are summarised in Table~\ref{table:observations}.

The calibration and imaging were done in CASA\footnote{\tt http://casa.nrao.edu/} \citep{McMullin2007}: the complex gains were calibrated through observations of the quasars J1517-2422 and J1625-2527, passband calibration on J1924-2914 and flux calibration on Titan. The resulting dataset has a beam size of 0\farcs44 $\times$ 0\farcs33, a continuum \textit{rms} level of 0.4~mJy~beam$^{-1}$ and a spectral \textit{rms} level of 10~mJy~beam$^{-1}$ per channel. The channel width can be 0.11 or 0.22~km~s$^{-1}$, depending on the spectral window (see Table~\ref{table:observations}).

\begin{table*}[t]
        \caption{Spectral setup and parameters of the detected molecular transitions.}
        \label{table:observations}
        \centering
        \begin{tabular}{l c c l c r c}
                \hline\hline
                Spectral window & Frequency range & Spectral resolution & Molecular transition & \textit{A$_{ij}$} $^{a}$ & \textit{E$_\mathrm{u}$} $^{a}$ & \textit{n$_\mathrm{crit}$} $^{b}$\\
                 & [GHz] & [km~s$^{-1}$] & &  [s$^{-1}$] &  [K] & [cm$^{-3}$] \\
                \hline
                spw 1a & 337.002$-$337.119 & 0.11 & C$^{17}$O \textit{J}=3$-$2 & 2.3 $\times$ 10$^{-6}$ & 32.3 & 3.5 $\times$ 10$^{4}$ \\
                spw 1b & 337.337$-$337.454 & 0.11 & C$^{34}$S \textit{J}=7$-$6 & 8.4 $\times$ 10$^{-4}$ & 50.2 & 1.3 $\times$ 10$^{7}$ \\
                spw 2 & 338.165$-$338.634 & 0.22  & c$-$C$_{3}$H$_{2}$ 5$_{5,1}$$-$4$_{4,0}$ (para) & 1.6 $\times$ 10$^{-3}$ & 48.8 & 1.2 $\times$ 10$^{8}$ \\
                 &  &  & SO$_{2}$ 18$_{4,14}$$-$18$_{3,15}$ & 3.3 $\times$ 10$^{-4}$ & 196.8 & 5.9 $\times$ 10$^{7}$ \\
                 &  &  & SO$_{2}$ 20$_{1,19}$$-$19$_{2,18}$ & 2.9 $\times$ 10$^{-4}$ & 198.9 & 3.8 $\times$ 10$^{7}$ \\
                spw 3 & 346.880$-$347.115 & 0.11 & H$^{13}$CO$^{+}$ \textit{J}=4$-$3 & 3.3 $\times$ 10$^{-3}$ & 41.6 & 8.5 $\times$ 10$^{6}$ \\
                spw 4 & 349.165$-$349.634 & 0.22 & c$-$C$_{3}$H$_{2}$ 5$_{5,0}$$-$4$_{4,1}$ (ortho) & 1.6 $\times$ 10$^{-3}$ & 49.0 & 1.3 $\times$ 10$^{8}$ \\
                 &  &  & C$_{2}$H \textit{N}=4$-$3, \textit{J}=9/2$-$7/2, \textit{F}=5$-$4 & 1.3 $\times$ 10$^{-4}$ & 41.9 & 2.2 $\times$ 10$^{7}$ \\
                 &  &  & C$_{2}$H \textit{N}=4$-$3, \textit{J}=9/2$-$7/2, \textit{F}=4$-$3 & 1.3 $\times$ 10$^{-4}$ & 41.9 & 2.3 $\times$ 10$^{7}$ \\
                 &  &  & C$_{2}$H \textit{N}=4$-$3, \textit{J}=7/2$-$5/2, \textit{F}=4$-$3 & 1.2 $\times$ 10$^{-4}$ & 41.9 & 1.7 $\times$ 10$^{7}$ \\
                 &  &  & C$_{2}$H \textit{N}=4$-$3, \textit{J}=7/2$-$5/2, \textit{F}=3$-$2 & 1.2 $\times$ 10$^{-4}$ & 41.9 & 1.8 $\times$ 10$^{7}$ \\
                \hline
        \end{tabular}
        \tablefoot{$^{(a)}$  Values from the CDMS database \citep{Muller2001}. $^{(b)}$ Calculated values for a collisional temperature of 30~K and collisional rates from the Leiden Atomic and Molecular Database \citep[LAMDA; ][]{Schoier2005}. The collisional rates of specific species were taken from the following works: C$^{17}$O from \cite{Yang2010}, C$^{34}$S from \cite{Lique2006}, c$-$C$_{3}$H$_{2}$ from \cite{Chandra2000}, SO$_{2}$ from \cite{Balanca2016}, H$^{13}$CO$^{+}$ from \cite{Flower1999} and C$_{2}$H from \cite{Spielfiedel2012}.}
\end{table*}

\section{Results}

\subsection{Continuum emission}

\begin{figure}[h]
\centering
   \includegraphics[width=.49\textwidth]{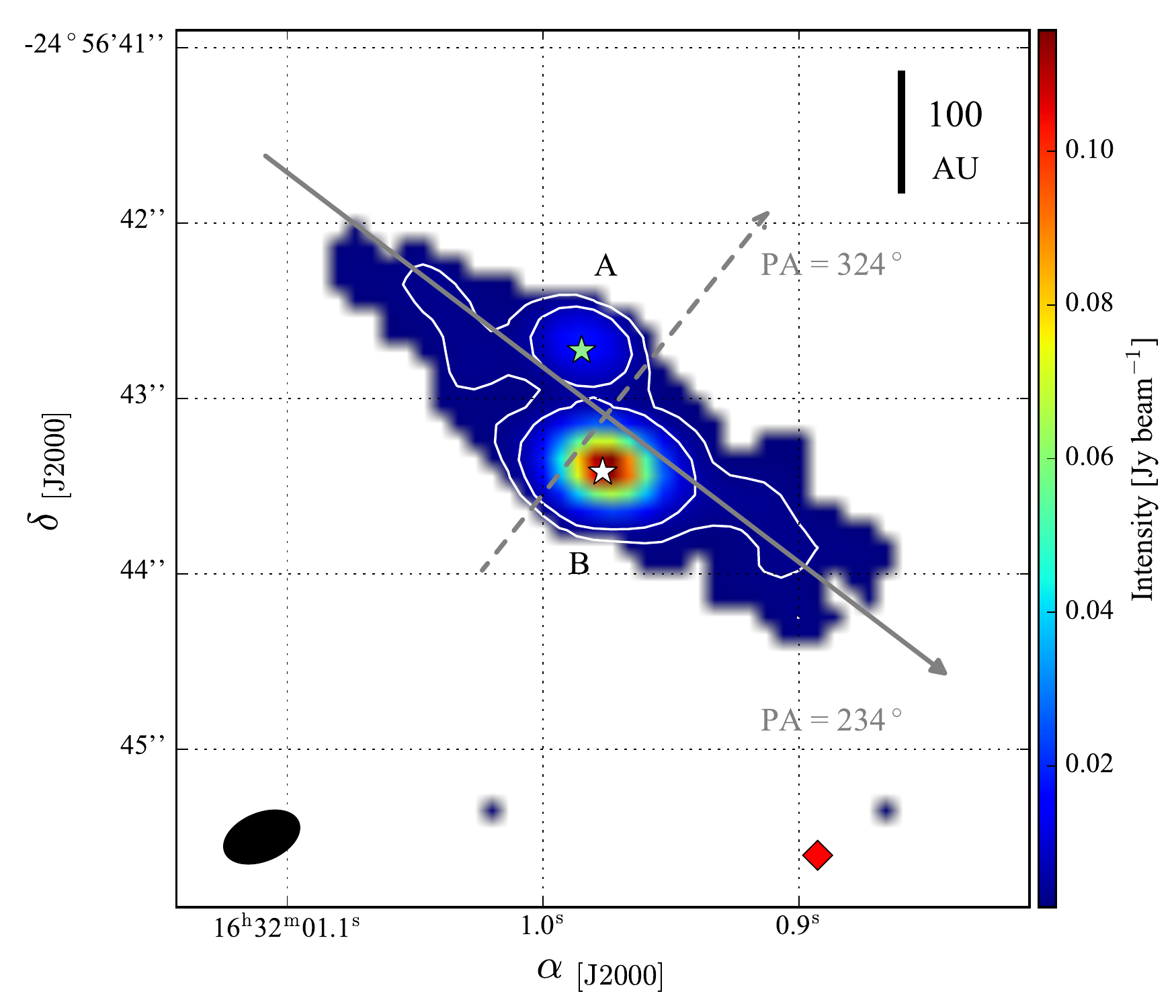}
      \caption{Continuum emission above 4$\sigma$ ($\sigma$ =  0.4~mJy~beam$^{-1}$) in colour scale and specific values of 7 and 15$\sigma$ in white contours. The synthesised beam is represented by the black filled ellipse. The grey solid and dashed arrows cross at the geometric centre and they represent the direction of and perpendicular to the disk-like structure, respectively. The green and white stars show the positions of Oph-IRS67A and Oph-IRS67B, respectively. The red diamond denotes the location of the offset region. 
      \label{fig:continuum}}
\end{figure}

Figure~\ref{fig:continuum} shows the submillimetre continuum emission toward Oph-IRS67, where two peaks are detected toward the A and B sources plus a fainter disk-like structure. Oph-IRS67B is brighter than Oph-IRS67A (155 vs. 30~mJy) and was identified by \citet{McClure2010} as a binary companion at infrared wavelengths. The disk-like structure has a deconvolved continuum size of (4\farcs1 $\pm$ 0\farcs2) $\times$ (0\farcs82 $\pm$ 0\farcs05) or (620 $\pm$ 20~AU) $\times$ (124 $\pm$ 7~AU), and a position angle (PA) of 54\degr $\pm$ 1\degr, measured from north to east. Assuming a toy model for a thin and circular structure, an inclination of $\sim$ 80$\degr$ (\textit{i} = 0$\degr$ for face-on and \textit{i} = 90$\degr$ for edge-on) is found by fixing the major and minor axis values of the deconvolved continuum size.

The position and integrated flux of each source are obtained by fitting two-dimensional (2D) Gaussians to both peaks and are listed in Table~\ref{table:coordinates}. The position of the geometric centre is also calculated, by taking the middle point between both sources. The separation between sources A and B is 0\farcs71 $\pm$ 0\farcs01 (107 $\pm$ 2~AU). All the figures and calculations in this work are shown relative to the geometric centre, and relative to the system velocity ($V_{\rm LSR}$) of 4.2 km~s$^{-1}$ \citep{Lindberg2017}. The red diamond in Fig.~\ref{fig:continuum} represents the position of an offset region, that may be associated with a high-density region, and is discussed in more detail in Sect. 3.2.1.

\begin{table*}[t]
\caption{Position and continuum fluxes of the system.}
\label{table:coordinates}
\centering
\begin{tabular}{l c c c}
        \hline\hline
                 & $\alpha$$_\mathrm{J2000}$ & $\delta$$_\mathrm{J2000}$ & Integrated flux \\
                 & & & [mJy] \\
        \hline
                Oph-IRS67A & 16$^\mathrm{h}$32$^\mathrm{m}$00.989$^\mathrm{s}$ $\pm$ 0.001$^\mathrm{s}$ & $-$24$^\circ$56$'$42.75$''$ $\pm$ 0.01$''$ & 30 $\pm$ 3 \\
                Oph-IRS67B & 16$^\mathrm{h}$32$^\mathrm{m}$00.978$^\mathrm{s}$ $\pm$ 0.001$^\mathrm{s}$ & $-$24$^\circ$56$'$43.44$''$ $\pm$ 0.01$''$ & 155 $\pm$ 4 \\
                Geometric center & 16$^\mathrm{h}$32$^\mathrm{m}$00.983$^\mathrm{s}$ $\pm$ 0.001$^\mathrm{s}$ & $-$24$^\circ$56$'$43.09$''$ $\pm$ 0.01$''$ & --- \\
        \hline
\end{tabular}
\end{table*}

\subsection{Molecular emission}

Table~\ref{table:observations} contains the parameters of the detected molecular transitions. The original proposal aimed at detecting the CH$_{3}$OH \textit{J}$_{k}$=7$_{k}$--6$_{k}$ branch, but no significant emission above 3$\sigma$ is observed. In the case of C$_{2}$H, the four lines compose two pairs of hyperfine splittings of two rotational levels. Since both hyperfine transitions in each pair are very close to one another, the C$_{2}$H transitions are labelled as \textit{N}=4$-$3, \textit{J}=9/2$-$7/2 and \textit{N}=4$-$3, \textit{J}=7/2$-$5/2. All the listed species are detected towards at least one of the following regions: source A, source B, the disk-like structure or a region spatially offset from the system ($\sim$430~AU from the geometrical center), located South-West from the continuum emission (red diamond in Fig.~\ref{fig:continuum}).

\subsubsection{Moment maps}    

\begin{figure*}[!]
   \centering
      \includegraphics[width=0.85\textwidth]{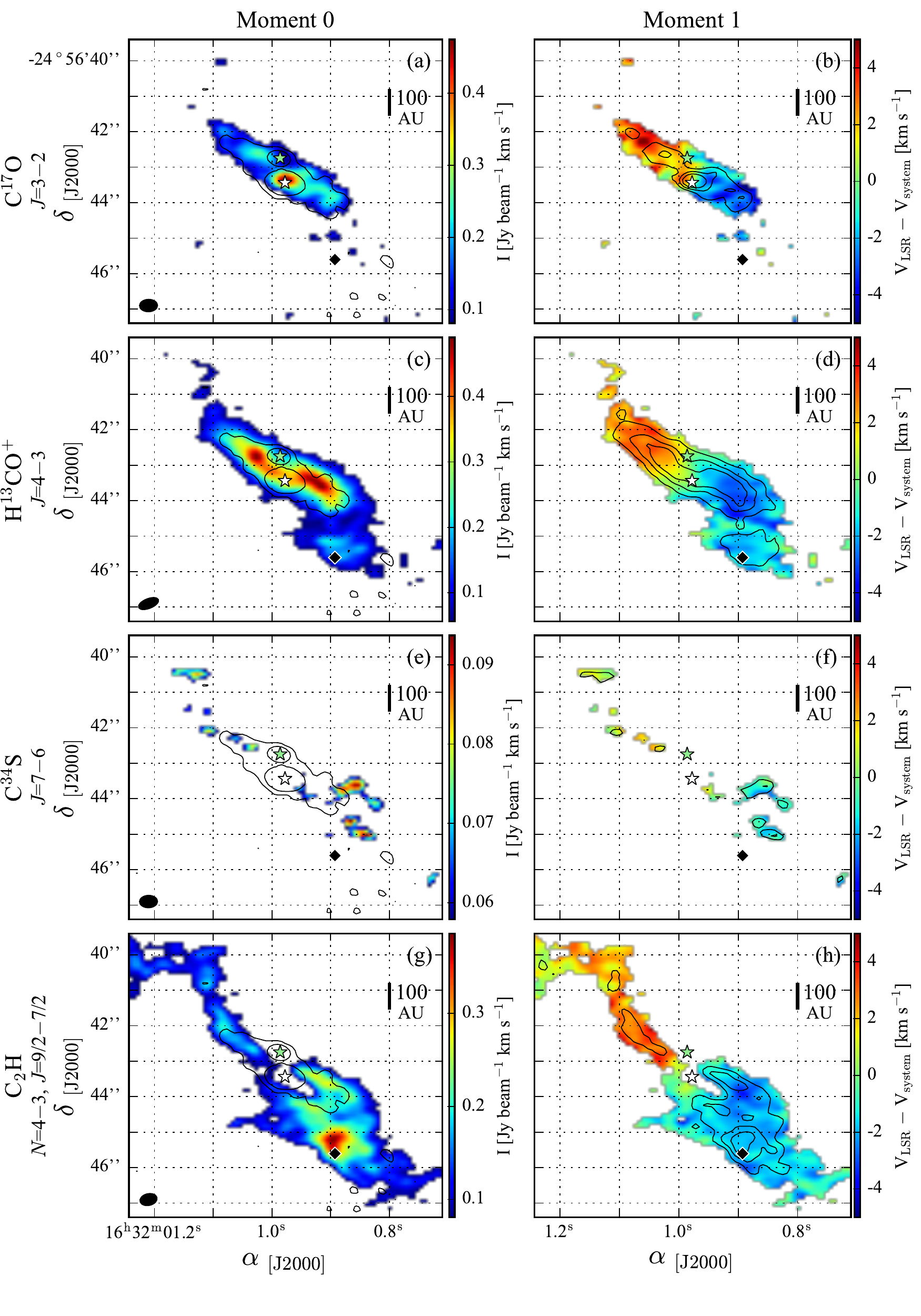}
      \caption[]{\label{fig:moments}
      Moments 0 (\textit{left column}) and 1 (\textit{right column}) for C$^{17}$O, H$^{13}$CO$^{+}$, C$^{34}$S and the C$_{2}$H \textit{N}=4$-$3, \textit{J}=9/2$-$7/2 doublet, above 5$\sigma$ ($\sigma$$_\mathrm{C^{17}O}$ =15~mJy~beam$^{-1}$~km~s$^{-1}$, $\sigma$$_\mathrm{H^{13}CO^{+}}$ = 10~mJy~beam$^{-1}$~km~s$^{-1}$, $\sigma$$_\mathrm{C^{34}S}$ = 10~mJy~beam$^{-1}$~km~s$^{-1}$ and $\sigma$$_\mathrm{C_{2}H}$ = 15~mJy~beam$^{-1}$~km~s$^{-1}$). Each $\sigma$ was calculated using the following formula: $\sigma$~=~\textit{rms}~$\times$~\textit{$\Delta$N}~$\times$~(\textit{N})$^{0.5}$, where \textit{$\Delta$N} and \textit{N} are the channel width and number of channels, respectively. Black contours in moment 0 maps represent the continuum emission from Fig.~\ref{fig:continuum}, for values of 4 and 15$\sigma$. Black contours in moment 1 maps show $\sigma$ values of their respective moment 0 maps, being 10, 15 and 20$\sigma$ for C$^{17}$O and C$_{2}$H, 10, 20 and 30$\sigma$ for H$^{13}$CO$^{+}$, and 6 and 8$\sigma$ for C$^{34}$S. The green and white stars indicate the position of Oph-IRS67A and B, respectively. The black diamond denotes the location of the offset region. The synthesised beam for each species is represented by a black filled ellipse.
      }
\end{figure*}

\begin{figure*}[!]
   \centering
      \includegraphics[width=0.85\textwidth]{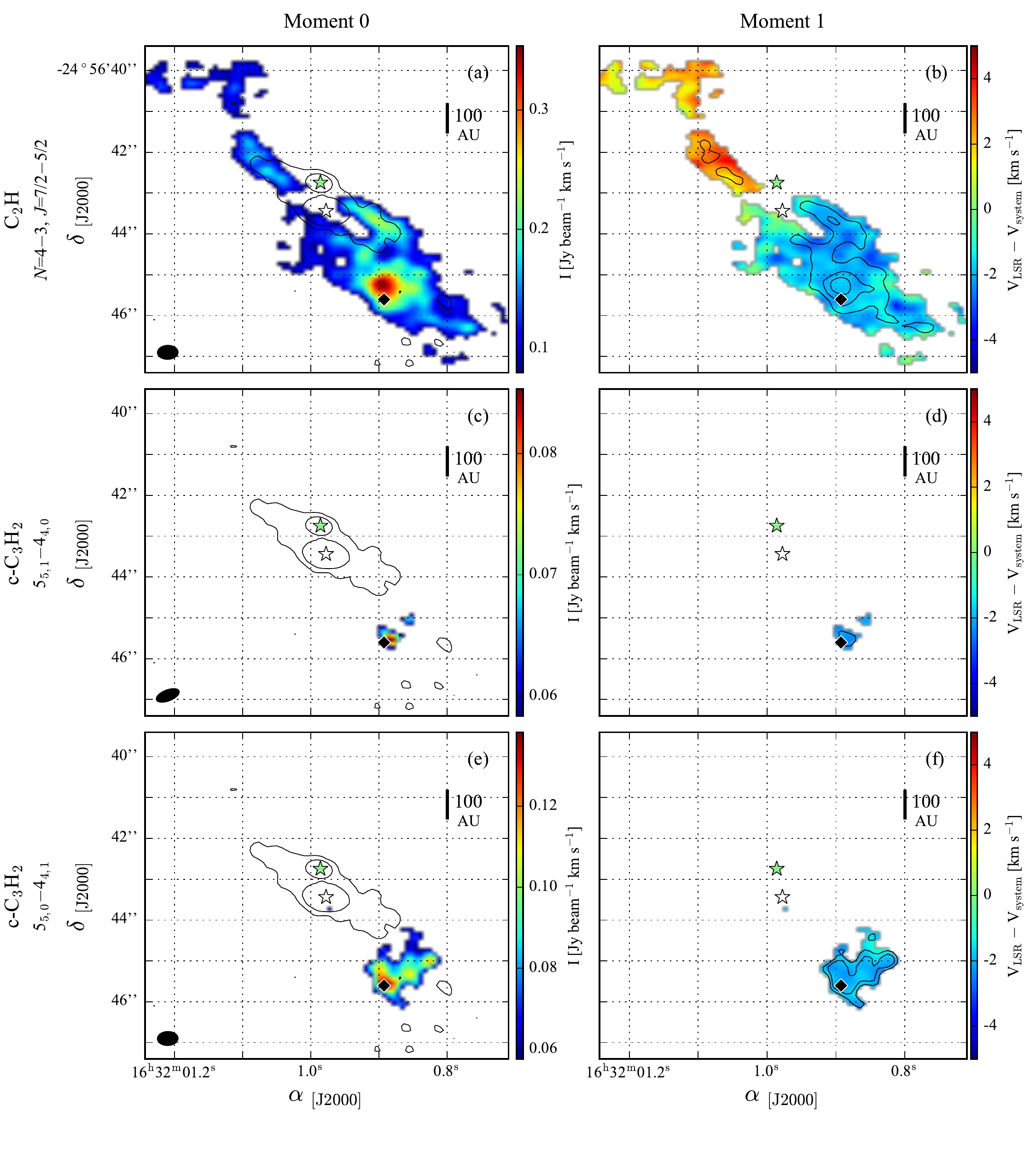}
      \includegraphics[width=0.87\textwidth]{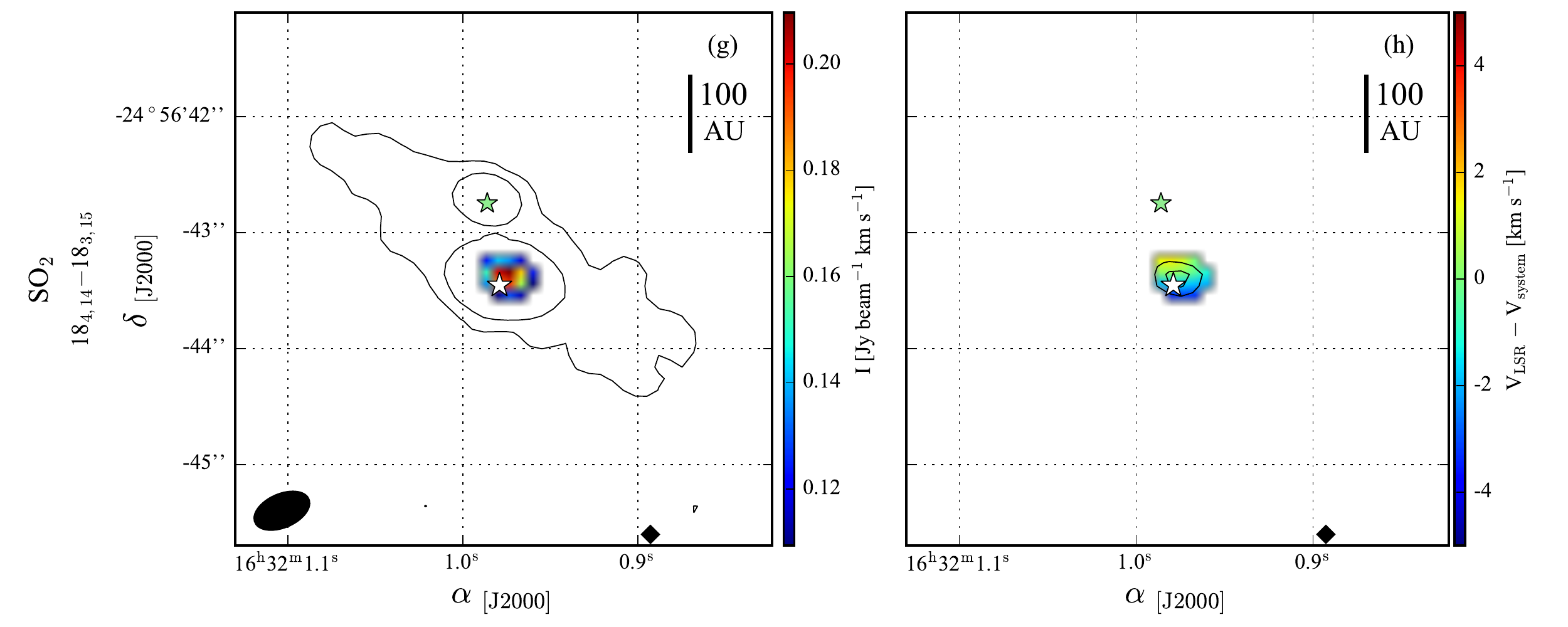}
      \caption[]{\label{fig:moments_bis}
     As in Fig.~\ref{fig:moments} but for the C$_{2}$H \textit{N}=4$-$3, \textit{J}=7/2$-$5/2 doublet, both c$-$C$_{3}$H$_{2}$ transitions and SO$_{2}$ 18$_{4,14}$$-$18$_{3,15}$ ($\sigma$$_\mathrm{c-C_{3}H_{2}}$ = 11~mJy~beam$^{-1}$~km~s$^{-1}$ and $\sigma$$_\mathrm{SO_{2}}$  = 25~mJy~beam$^{-1}$~km~s$^{-1}$).  For SO$_{2}$, the emission is above 4$\sigma$. Black contours in moment 1 maps represent 10, 15 and 20$\sigma$ for C$_{2}$H, 6 and 8$\sigma$ for c$-$C$_{3}$H$_{2}$, and 5 and 7$\sigma$ for SO$_{2}$. Panels \emph{g} and \emph{h} represent a zoomed-in region. 
      }
\end{figure*}

Figures~\ref{fig:moments} and \ref{fig:moments_bis} present the integrated emission (moment 0) and the velocity field (moment 1) for C$^{17}$O, H$^{13}$CO$^{+}$, C$^{34}$S, the two doublets of C$_{2}$H, both transitions of c$-$C$_{3}$H$_{2}$, and SO$_{2}$. In moment 0 maps, the continuum emission from Fig.~\ref{fig:continuum} corresponding to 4 and 15$\sigma$ contours is also plotted. C$^{17}$O is tracing the disk-like structure, with a strong correlation with the continuum emission and its integrated intensity peaks at the position of source B. The H$^{13}$CO$^{+}$ emission is very intense in the proximity of both sources, showing an S-shape correlated with the continuum emission, and is enhanced in the offset region. C$^{34}$S peaks in isolated regions away from both sources and part of the emission lies beyond the continuum emission. The lines of C$_{2}$H display the most extended emission and their integrated intensity dominates in the offset region. The North-East emission presents a curved shape beyond the continuum emission, while the South-West region is almost completely dominated by C$_{2}$H emission. Close to the protostars, the C$_{2}$H integrated intensity is below 5$\sigma$. Both c$-$C$_{3}$H$_{2}$ transitions show extended emission only in the vicinity of the offset region and are anti-correlated with the continuum emission. SO$_{2}$ shows compact emission only toward source B, where the continuum emission peaks. The velocity ranges in Figs.~\ref{fig:moments} and \ref{fig:moments_bis} are chosen to bring forward the disk-like structure. For a detailed comparison of the more quiescent gas associated with the individual protostars, see the channel maps in Figs. A.1, A.2 and A.3 in Appendix A.

The offset region shows strong emission of the two c$-$C$_{3}$H$_{2}$ transitions. Since these two transitions are associated with high critical densities (\textit{n$_\mathrm{crit}$} = 1.2 $\times$ 10$^{8}$ and 1.4 $\times$ 10$^{8}$ cm$^{-3}$; see Table~\ref{table:observations}), the offset region is henceforth referred to as the \textit{high density region}.
 
The moment 1 maps in Figs.~\ref{fig:moments} and \ref{fig:moments_bis} show the existence of a velocity gradient and suggest an approximately edge-on and flat morphology for the disk-like structure, with the blue-shifted emission arising from the South-West region and the red-shifted emission emerging from the North-East area. In addition, the high-density region only shows blue-shifted emission with no red-shifted counterpart, and its emission peaks around 2~km~s$^{-1}$. SO$_{2}$ emission shows a velocity gradient toward source B (PA~$\approx$~30$\degr$), misaligned with the gradient seen for the other detected lines (PA~$\approx$~54\degr). This orientation difference may reveal that a possible circumstellar disk around source B is misaligned with respect to the disk-like structure, if SO$_{2}$ is tracing circumstellar disk material, or SO$_{2}$ may be tracing accretion shocks and thus, related with infalling motion.

\subsubsection{Spectra}    
  
\begin{figure*}[!]
   \centering
      \includegraphics[width=.85\textwidth]{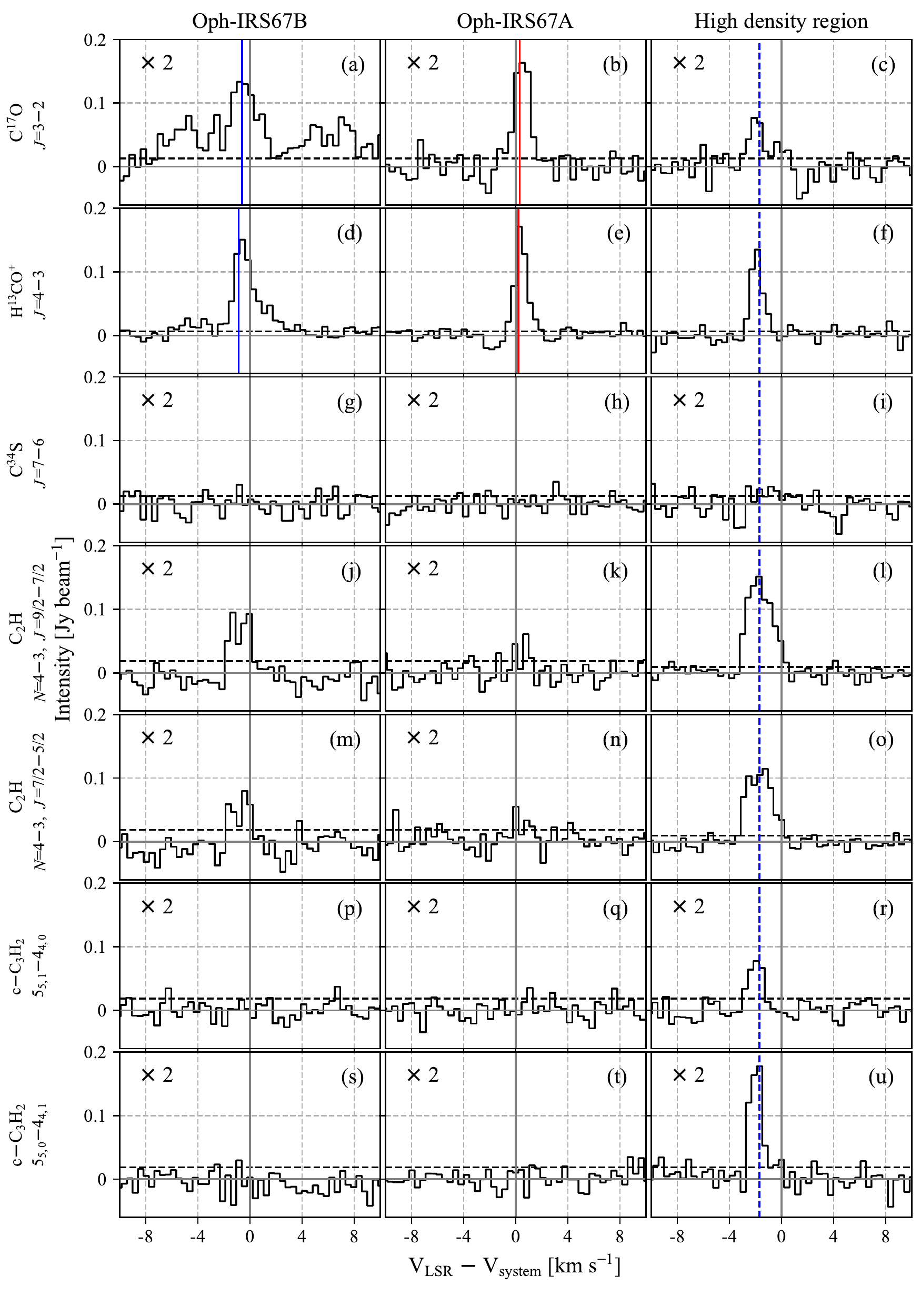}
      \caption[]{\label{fig:spectra}
      Spectra of C$^{17}$O, H$^{13}$CO$^{+}$, C$^{34}$S, both C$_{2}$H doublets and both c$-$C$_{3}$H$_{2}$ transitions for three different regions: the position of Oph-IRS67B (\textit{left column}), the position of Oph-IRS67A (\textit{middle column}) and the high-density region (\textit{right column}). Spectra from panels \textit{(a)} to \textit{(i)} are rebinned spectrally by a factor of 4, while panels from \textit{(j)} to \textit{(u)} are rebinned spectrally by a factor of 2, so that all spectra have the same spectral resolution (0.43~km~s$^{-1}$). The dashed black horizontal line shows the value of 3$\sigma$ ($\sigma$ = 3~mJy~beam$^{-1}$~km~s$^{-1}$). All spectra have been shifted to the systemic velocity (4.2~km~s$^{-1}$; grey vertical line). The solid blue and red lines indicate the offset of the peak velocity determined from a Gaussian fit (see Table~\ref{table:Gfit}). The dashed blue line in the right column represents a velocity of $-$1.7~km~s$^{-1}$. Some spectra have been multiplied by a scaling factor as indicated in the top left corner.
      }
\end{figure*}

Figure~\ref{fig:spectra} shows the spectra of C$^{17}$O, H$^{13}$CO$^{+}$, C$^{34}$S, C$_{2}$H and both c$-$C$_{3}$H$_{2}$ transitions towards the positions of source B, source A, and the high density region. Each spectrum was extracted over 1 pixel, that is, 0\farcs1~$\times$~0\farcs1. At the position of source B, C$^{17}$O shows a central component with two clear peaks at higher velocities (between $-$8 and $-$4~km~s$^{-1}$ and $+$4 and $+$8~km~s$^{-1}$), H$^{13}$CO$^{+}$ peaks near $V_{\rm LSR}$ with a wing morphology that extends over positive velocities, and both C$_{2}$H doublets present two peaks, related with each hyperfine transition. C$^{34}$S and both c$-$C$_{3}$H$_{2}$ transitions are not detected towards source B. Source A is associated with narrower lines from C$^{17}$O and H$^{13}$CO$^{+}$, showing only a central component, and a weaker contribution from C$_{2}$H. As for source B, C$^{34}$S and c$-$C$_{3}$H$_{2}$ are not detected towards source A. The high-density region presents weaker emission of C$^{17}$O and H$^{13}$CO$^{+}$, where both lines peak at $-$1.7~km~s$^{-1}$. The two C$_{2}$H doublets show intense and broader lines, and both c$-$C$_{3}$H$_{2}$ transitions are enhanced in this region. Observing the emission of C$^{17}$O and H$^{13}$CO$^{+}$ over both sources, the peak in the spectra associated with Oph-IRS67B appears slightly blue-shifted (Fig.~\ref{fig:spectra}a and d) with respect to $V_{\rm LSR}$, while Oph-IRS67A shows a red-shifted displacement (Fig.~\ref{fig:spectra}b and e).

\begin{figure}[t]
   \centering
      \includegraphics[width=.49\textwidth]{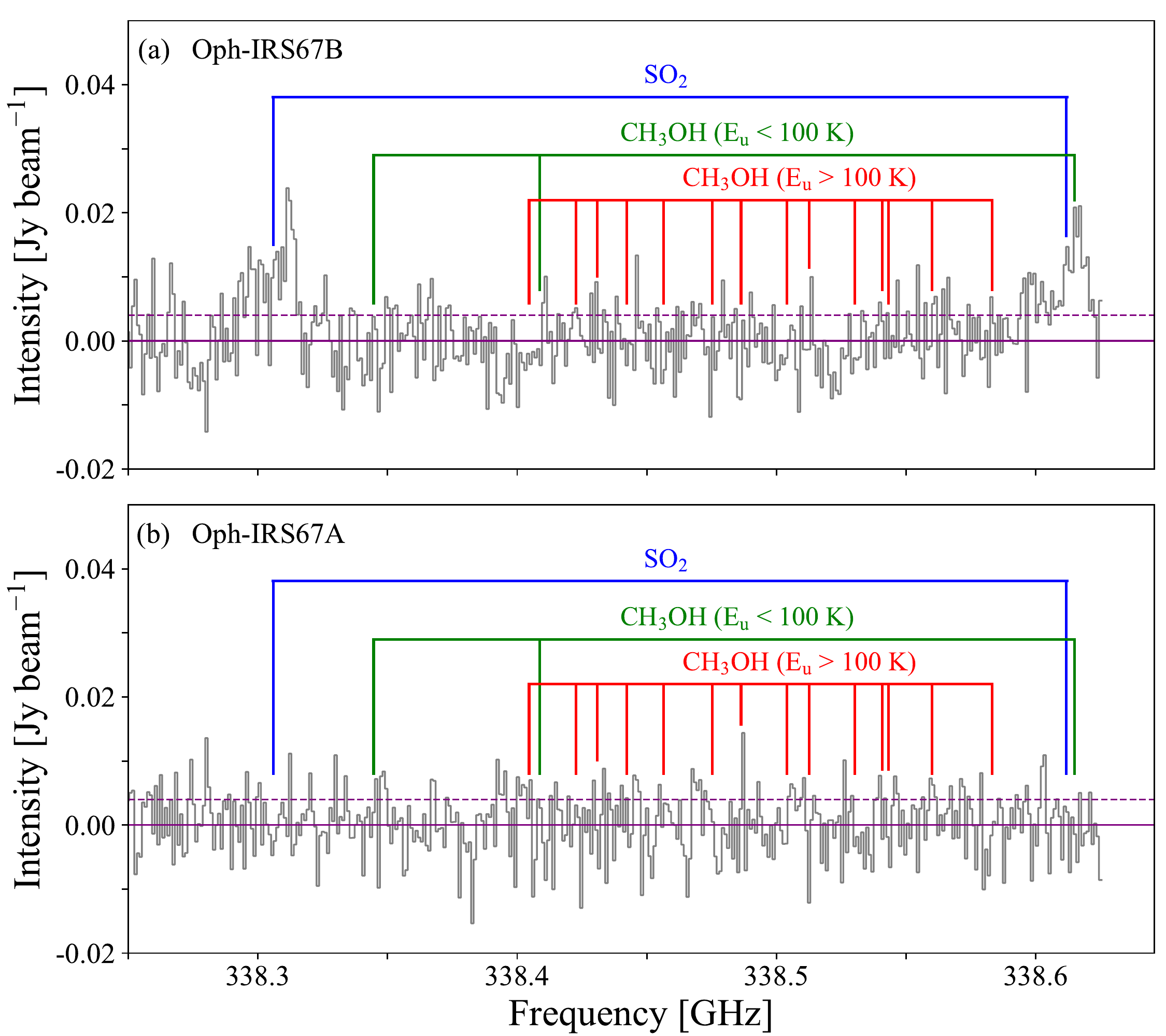}
      \caption[]{\label{fig:spectra_SO2}
      Spectra from spectral window number 2, taken at the position of Oph-IRS67B (\textit{top}) and Oph-IRS67A (\textit{bottom}), and rebinned by a factor of 4. Blue lines show the rest frequency of SO$_{2}$ transitions. Green and red lines indicate the rest frequency of CH$_{3}$OH transitions with upper energy levels (\textit{E$_{u}$}) below and above 100~K, respectively. The purple dashed line shows the value of 1$\sigma$ ($\sigma$ = 4~mJy~beam$^{-1}$ per channel).
      }
\end{figure}

Figure~\ref{fig:spectra_SO2} shows part of the spectral window number 2 toward the position of source B and source A, where the rest frequencies of SO$_{2}$ and CH$_{3}$OH are indicated. SO$_{2}$ transitions are only seen towards source B and show broad emission (from $-$8 to 10~km~s$^{-1}$), with an intense blue-shifted component and a weaker red-shifted one. One of the CH$_{3}$OH \textit{J$_{k}$}=$7_{k}$--$6_{k}$ transitions falls close to a SO$_{2}$ line (Fig.~\ref{fig:spectra_SO2}a), but since no other lines in the CH$_{3}$OH branch are seen, including the lower excitation transitions, the line can clearly be attributed to SO$_{2}$.

\subsubsection{Channel maps}    

\begin{figure*}[t]
   \centering
      \includegraphics[width=0.98\textwidth]{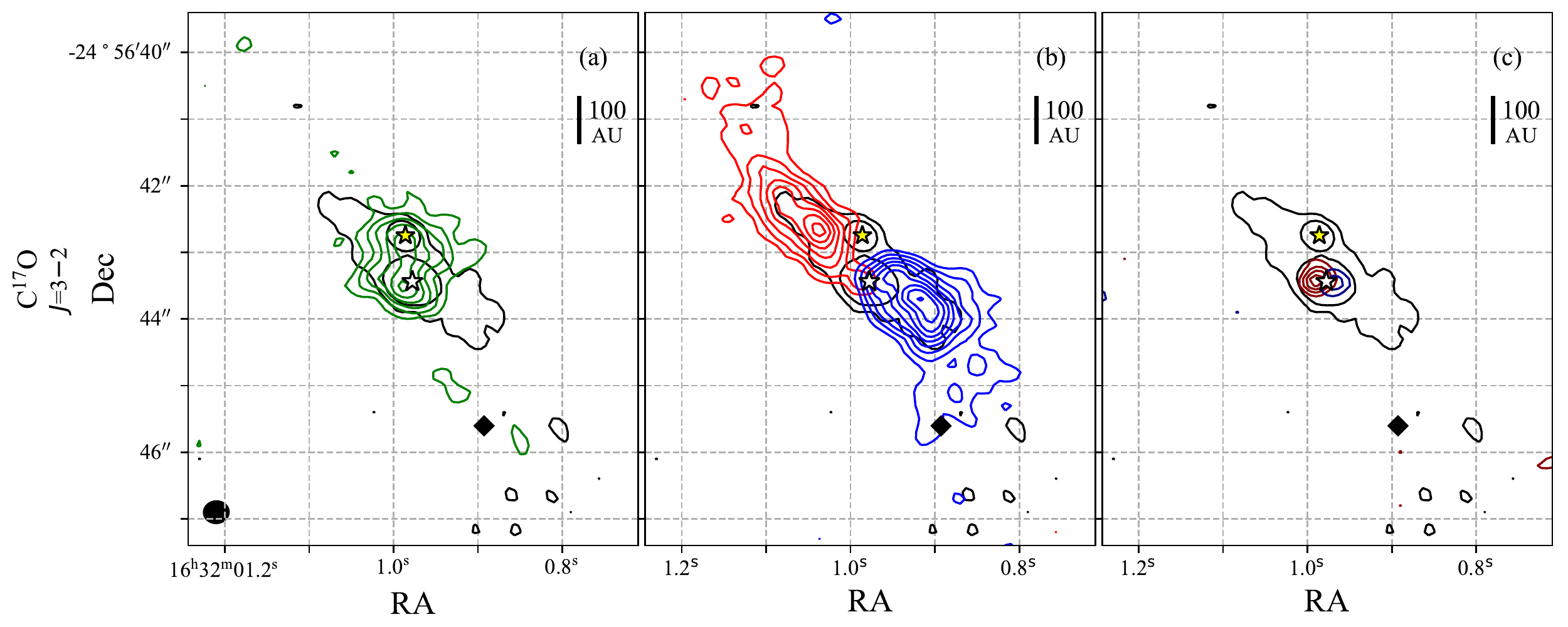}
      \caption[]{\label{fig:channels_CO}
      Channel maps for C$^{17}$O, starting at 5$\sigma$ ($\sigma$ = 6~mJy~beam$^{-1}$~km~s$^{-1}$) and following a step of 4$\sigma$. Green contours represent low velocities emission, from $-$1 to 1~km~s$^{-1}$ (\textit{left}). Blue and red contours indicate intermediate velocities, from $-$1 to $-$5~km~s$^{-1}$ and from 1 to 5~km~s$^{-1}$, respectively (\textit{middle}). Dark-red and dark-blue contours show high-velocity emission, from $-$5 to $-$10~km~s$^{-1}$ and from 5 to 10~km~s$^{-1}$, respectively (\textit{right}). The black contours represent the continuum emission of Fig.~\ref{fig:continuum} for values of 4 and 15$\sigma$ ($\sigma$~=~0.4mJy~beam$^{-1}$). The yellow and white stars indicate the position of Oph-IRS67A and B, respectively. The black diamond denotes the location of the high density region. The synthesised beam is represented by a black filled ellipse.
      }
\end{figure*}

\begin{figure*}[t]
   \centering
      \includegraphics[width=.49\textwidth]{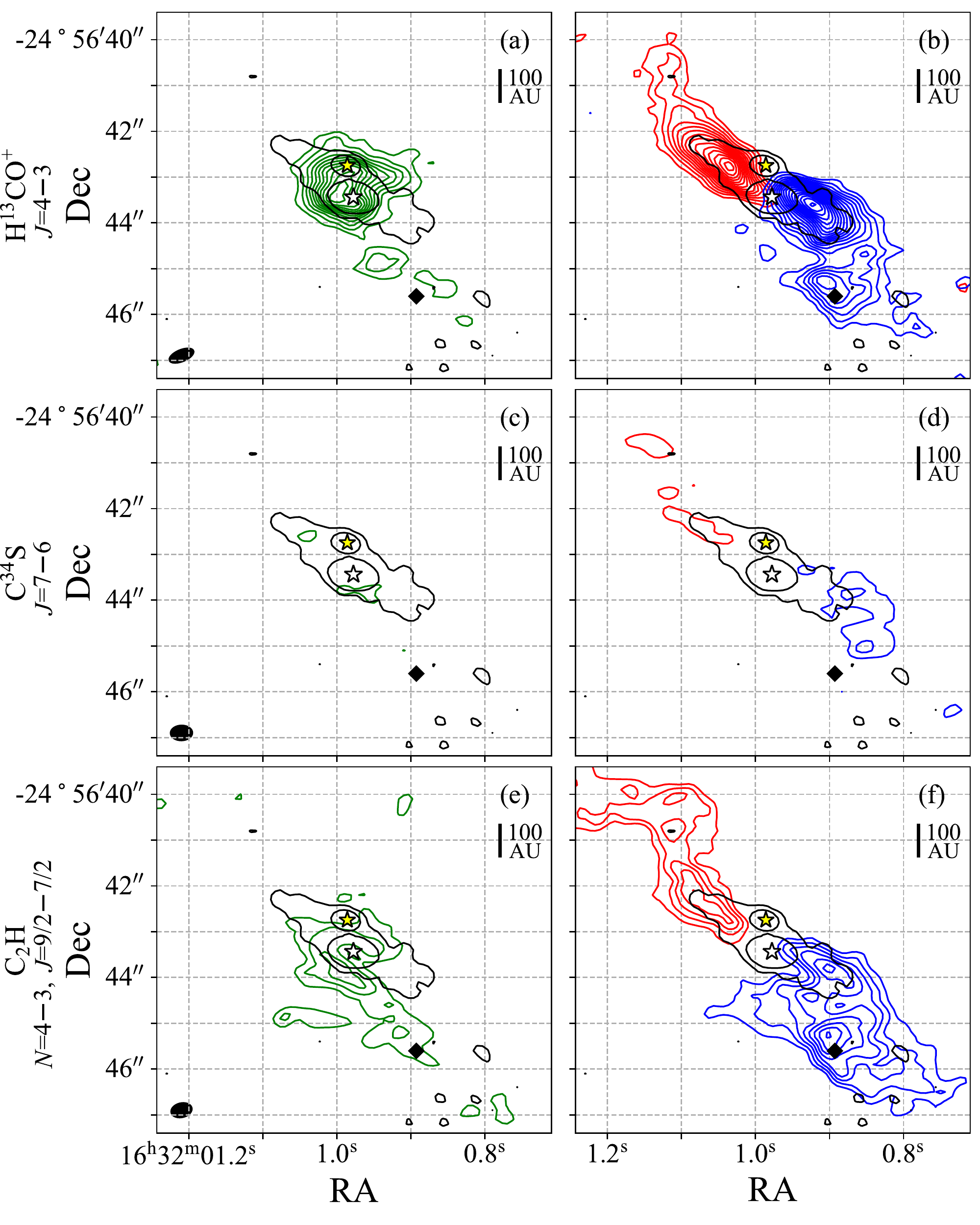}
      \includegraphics[width=.49\textwidth]{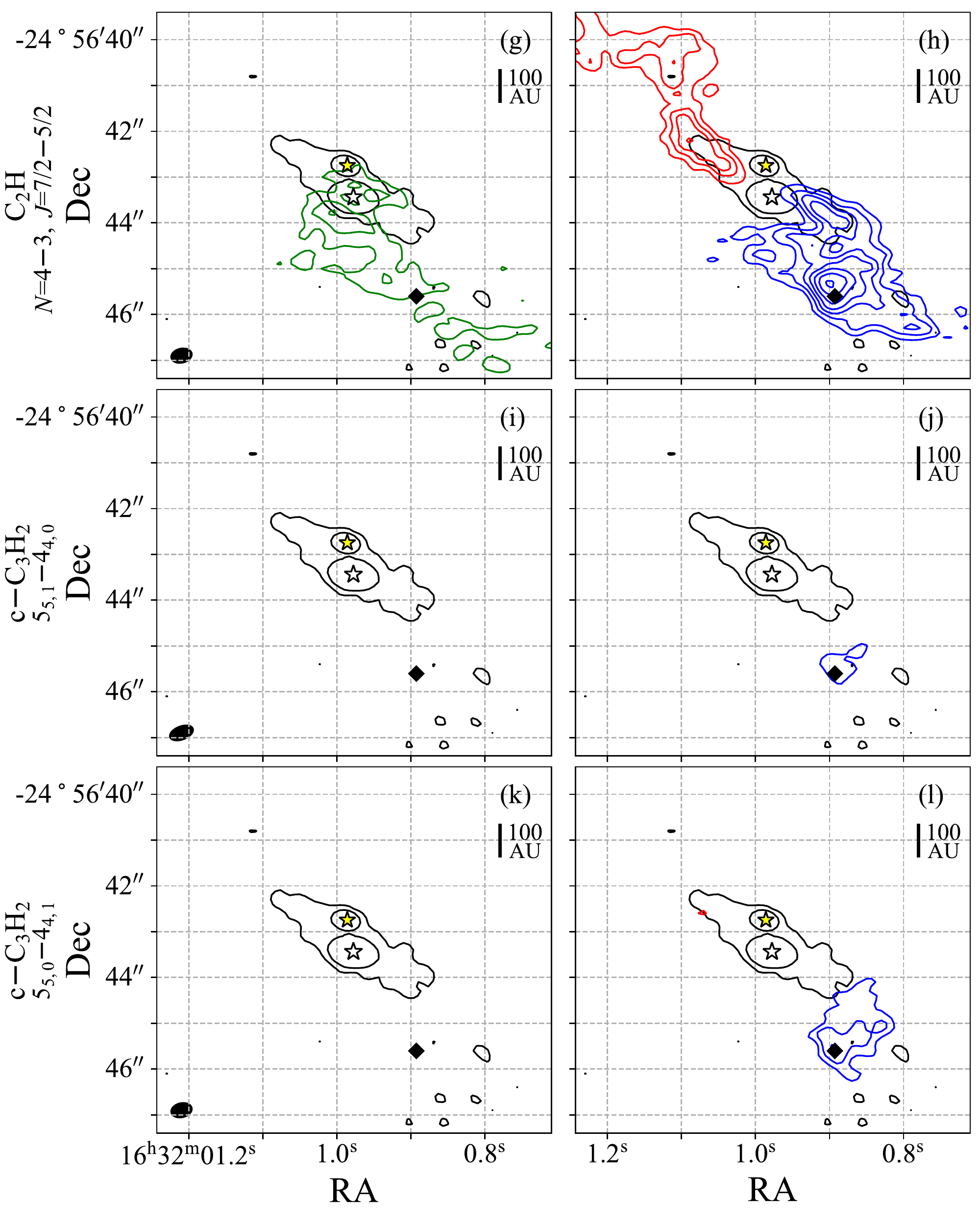}
      \caption[]{\label{fig:channels}
      Contour maps for H$^{13}$CO$^{+}$, C$^{34}$S, both C$_{2}$H doublets and both c$-$C$_{3}$H$_{2}$ transitions, starting at 5$\sigma$ and following a step of 4$\sigma$ ($\sigma$$_\mathrm{H^{13}CO^{+}}$ and $\sigma$$_\mathrm{C^{34}S}$ are 6~mJy~beam$^{-1}$~km~s$^{-1}$, while $\sigma$$_\mathrm{C_{2}H}$ and $\sigma$$_\mathrm{c-C_{3}H_{2}}$ are 9~mJy~beam$^{-1}$~km~s$^{-1}$). Green contours represent low-velocity emission, from $-$1 to 1~km~s$^{-1}$. Blue and red contours indicate intermediate velocities, from $-$1 to $-$5~km~s$^{-1}$ and from 1 to 5~km~s$^{-1}$, respectively. The black contours represent the continuum emission of Fig.~\ref{fig:continuum} for a value of 4 and 15$\sigma$ ($\sigma$~=~0.4mJy~beam$^{-1}$). The yellow and white stars indicate the position of Oph-IRS67A and B, respectively. The black diamond denotes the location of the high-density region. The synthesised beam for each species is represented by a black filled ellipse.
      }
\end{figure*}

\begin{figure}[t]
   \centering
      \includegraphics[width=.49\textwidth]{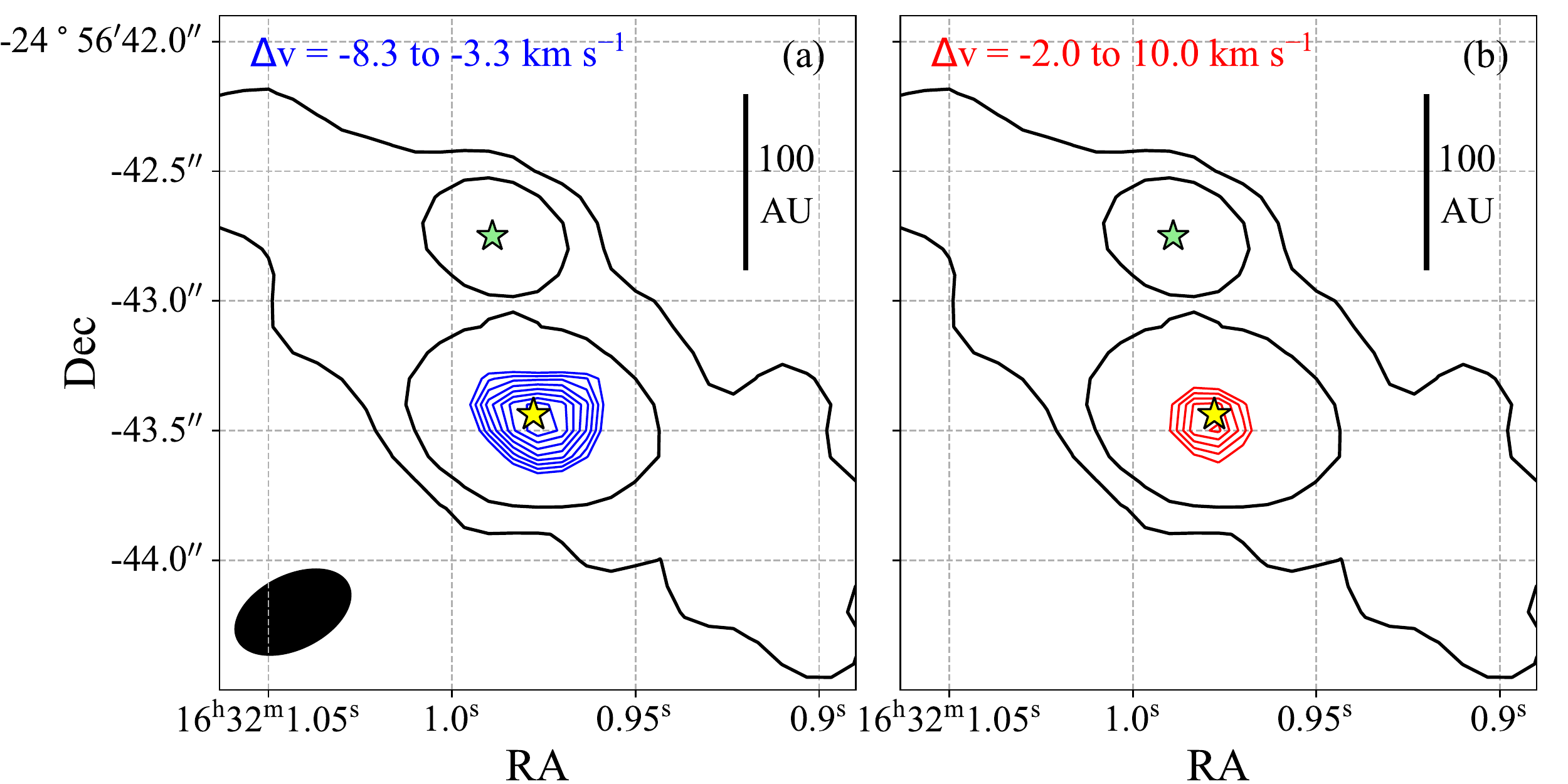}
      \caption[]{\label{fig:channels_SO2}
      Contour maps for SO$_{2}$ 18$_{4,14}$$-$18$_{3,15}$ for two ranges of velocities. The contours start at 4$\sigma$, following a step of 0.5$\sigma$. The black contours represent the continuum emission of Fig.~\ref{fig:continuum} for a value of 4 and 15$\sigma$ ($\sigma$~=~0.4mJy~beam$^{-1}$). The green and white stars indicate the position of Oph-IRS67A and B, respectively. The synthesised beam is represented by a black filled ellipse.
      }
\end{figure}

Contour maps for C$^{17}$O integrated over five different velocity ranges are shown in Fig.~\ref{fig:channels_CO}. The panels are divided into low velocities ($-$1 to 1~km~s$^{-1}$), intermediate velocities ($-$5 to $-$1~km~s$^{-1}$ and 1 to 5~km~s$^{-1}$) and high velocities ($-$10 to $-$5~km~s$^{-1}$ and 5 to 10~km~s$^{-1}$). H$^{13}$CO$^{+}$, C$^{34}$S, C$_{2}$H and c$-$C$_{3}$H$_{2}$ only show intermediate- and low-velocity components, thus, contour maps associated with these species are taken for three ranges of velocities (Fig.~\ref{fig:channels}). SO$_{2}$ shows a broad and double-peaked spectrum associated with blue- and red-shifted components (see Fig.~\ref{fig:spectra_SO2}a), therefore, the spectrum is fitted by a two-component Gaussian and the velocity ranges used for the channel maps of Fig.~\ref{fig:channels_SO2} are associated with the FWHM of the blue- and red-shifted components.

At low velocities, C$^{17}$O and H$^{13}$CO$^{+}$ appear to be centred around both sources. C$^{34}$S presents isolated peaks closer to source B, while C$_{2}$H dominates the central and Southern regions. A lack of emission at low velocities is seen for both c$-$C$_{3}$H$_{2}$ transitions.

At intermediate velocities, C$^{17}$O and H$^{13}$CO$^{+}$ show some symmetry with respect to Oph-IRS67B. C$^{34}$S presents an S-type morphology and enhanced emission far away from the binary system, while C$_{2}$H shows the most extended emission and the blue-shifted emission dominates over a wider region than the red-shifted one. Both c$-$C$_{3}$H$_{2}$ transitions show the same behaviour, with intermediate blue-shifted velocities tracing the high-density region and no red-shifted counterpart. 

High velocities are associated with C$^{17}$O and SO$_{2}$, showing compact emission only around Oph-IRS67B. C$^{17}$O blue- and red-shifted emission are strongly symmetric around the source, while the blue component associated with SO$_{2}$ stands out over the red one and is concentrated around the source.

\section{Analysis}
\subsection{Velocity profiles} 

C$^{17}$O and H$^{13}$CO$^{+}$ are the best candidates for studying the velocity profile of the disk-like structure through position-velocity (PV) diagrams, since they are correlated with the continuum emission. The other species are not suitable for studying the velocity profile of the disk-like structure, as they seem to be tracing different material. C$_{2}$H has an important contribution beyond 2$''$ and may be tracing envelope material, while C$^{34}$S is tracing smaller regions from the outer parts of the disk-like structure, and SO$_{2}$ emission is unresolved within the beam size. On the other hand, c$-$C$_{3}$H$_{2}$ is tracing the high-density region; its velocity profile is discussed in Sect. 4.1.2.

In order to obtain the velocity profiles, the peak emission for each channel is obtained through the CASA task \texttt{imfit}. Two fits are considered for high-velocity points (>1.7~km~s$^{-1}$): \textit{(i)} a Keplerian fit where \textit{$\varv$} $\propto$ \textit{r}$^{-0.5}$ and \textit{(ii)} an infalling fit where \textit{$\varv$} $\propto$ \textit{r}$^{-1}$ (i.e. infalling motion where the angular momentum is conserved). In all cases, the central (0,0) position corresponds to the location of the geometric centre and the distance increase toward the disk-like structure direction (PA = 234\degr; grey solid arrow in Fig.~\ref{fig:continuum}), assuming an inclination of 80$\degr$. 

\subsubsection{Disk-like structure} 

\begin{figure*}[t]
   \centering
      \includegraphics[width=.45\textwidth]{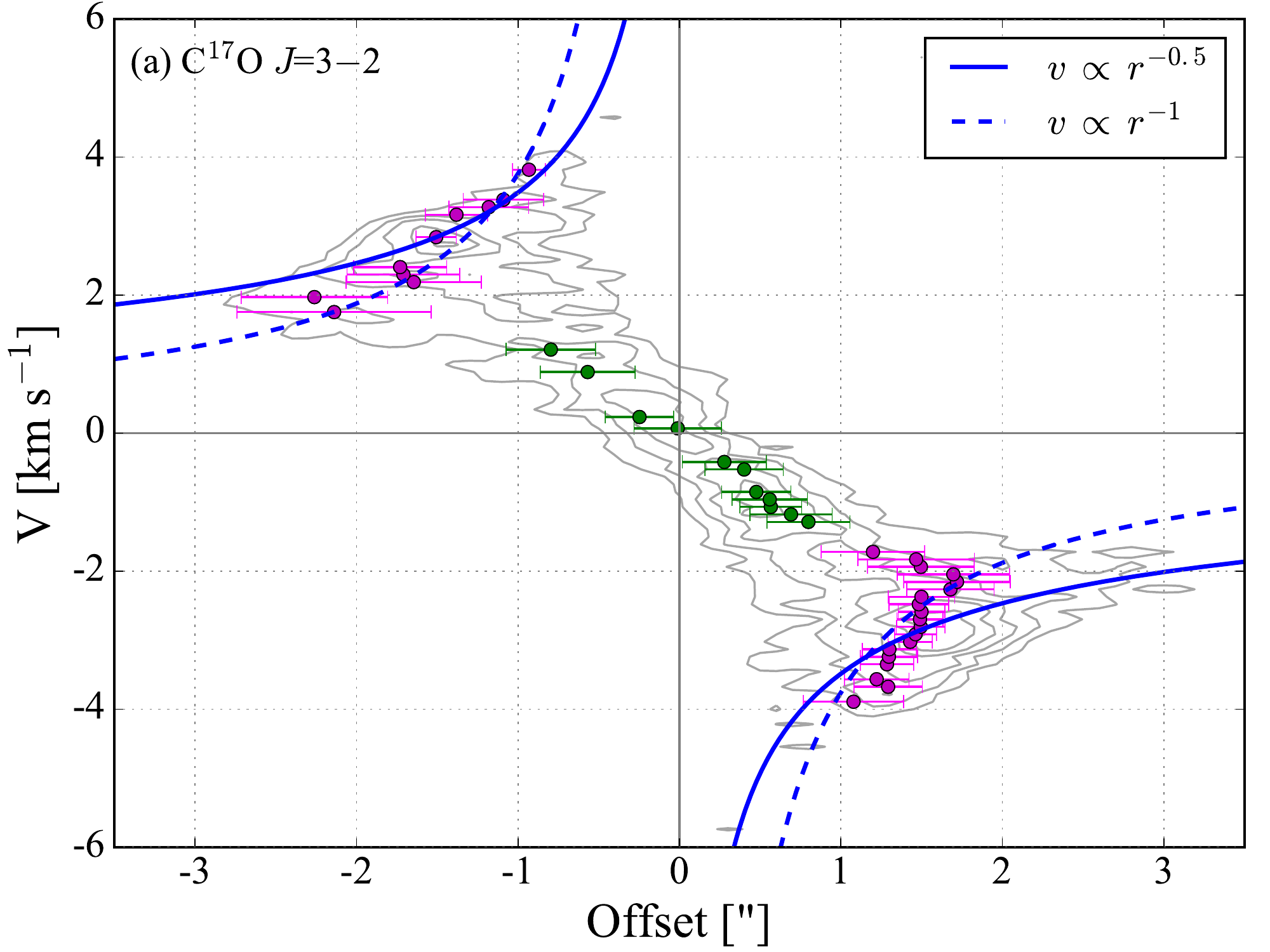}
      \includegraphics[width=.45\textwidth]{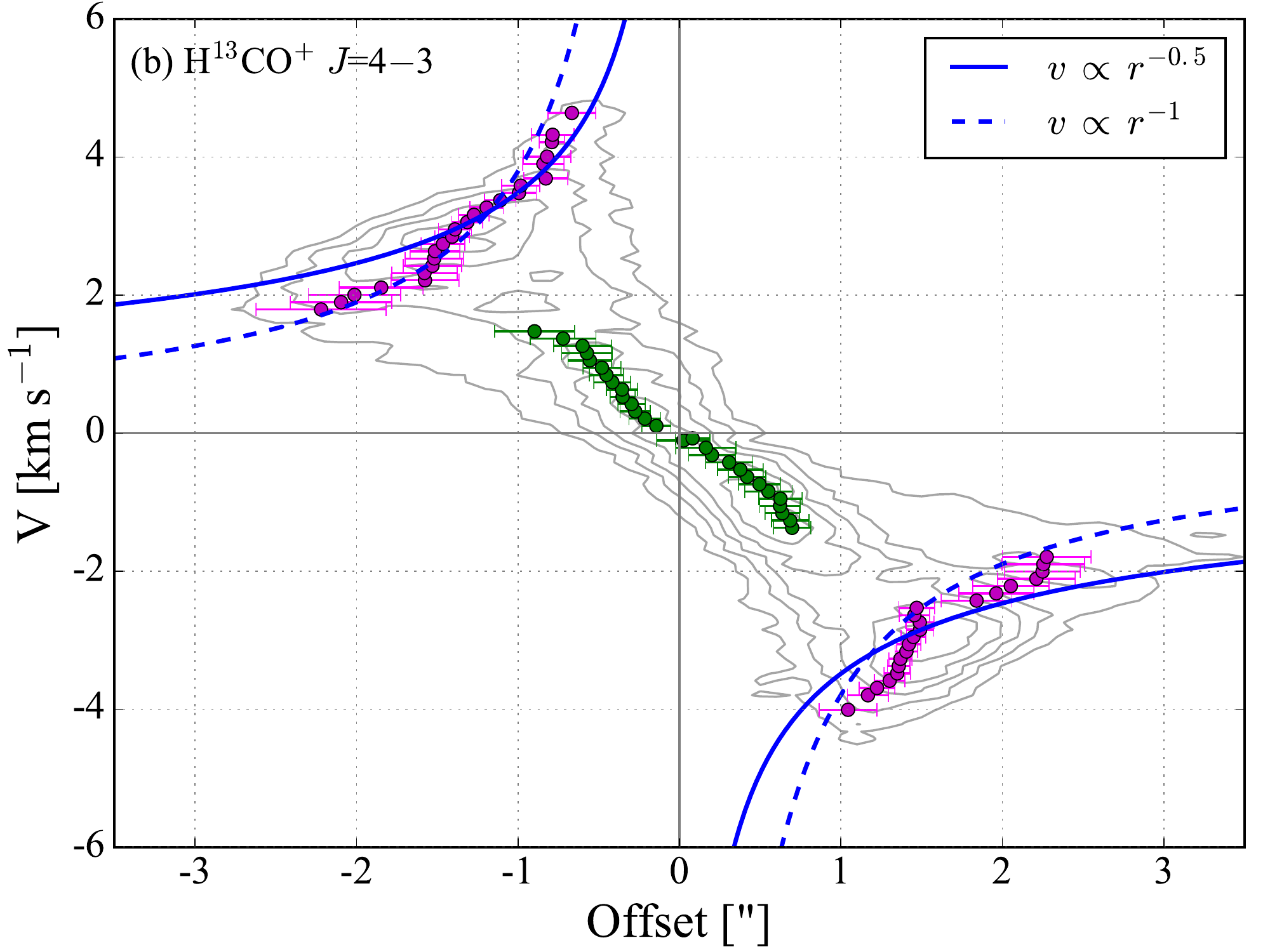}
      \caption[]{\label{fig:PIV}
          Position-velocity diagrams for C$^{17}$O and H$^{13}$CO$^{+}$ towards the disk-like structure direction (PA = 234$\degr$) and centred at the position of the geometric centre. Green and magenta dots represent low- (< 1.7~km~s$^{-1}$) and high-velocity (> 1.7~km~s$^{-1}$) emission peaks, respectively. Blue solid and dashed lines show the best fit for a Keplerian and an infalling velocity profile, respectively. The cut taken from the image data is shown in grey contours, ranging from 3$\sigma$ (1$\sigma$ = 10~mJy~beam$^{-1}$) to the maximum value of each transition. Each adjacent contour represents an increment of the 20$\%$ of the maximum value.
       }
\end{figure*}

The PV diagrams for C$^{17}$O and H$^{13}$CO$^{+}$ are shown in Fig.~\ref{fig:PIV}. Neither the Keplerian nor the infalling fit is representative of the data, where the points with velocities greater than 3~km~s$^{-1}$ seem to follow the Keplerian curve, while points with velocities between $\sim$~2 and 3~km~s$^{-1}$ tend to follow the infalling curve. This fact is highlighted by their reduced \textit{$\chi^{2}$} values, listed in Table~\ref{table:PIV}. For C$^{17}$O both fits are associated with the same \textit{$\chi^{2}_\mathrm{red}$}, while for H$^{13}$CO$^{+}$ the Keplerian fit has a lower \textit{$\chi^{2}_\mathrm{red}$} value than the infalling one. This means that the goodness of the Keplerian fit is better than the infalling fit, but neither of them can be discarded by this PV analysis. However, the presence of low-velocity points (< 1.7~km~s$^{-1}$) that appear to follow a linear distribution is expected for a Keplerian disk with finite radius \citep{Lindberg2014}. This emission arises from the edge of the disk and is an unlikely effect in the infall scenario, since the infalling material should not have a sharp edge.

The Keplerian and infalling fits were also tested for the scenario where the position of source B is used as the centre, resulting in \textit{$\chi^{2}_\mathrm{red}$} values higher than 10 for both fits. Therefore, the material from the disk-like structure is not moving with respect to either source, but with respect to some position between the two.

The PV diagrams taken towards the direction perpendicular to the disk-like structure (PA = 324\degr; grey dashed arrow in Fig.~\ref{fig:continuum}) did not show significant emission beyond 1$''$ (150~AU) and no infall signatures were detected. This lack of emission towards a perpendicular direction is consistent with Keplerian rotation around the binary system at small scales.

The presence of a linear velocity profile for low velocities in Fig.~\ref{fig:PIV} and the lack of infalling signatures towards a perpendicular direction suggests that \textit{circumbinary disk} is a reasonable label for the disk-like structure. Based on the Keplerian fit, the circumbinary disk is associated with a central mass of 2.2 $\pm$ 0.2~M$_{\odot}$.

\begin{table}[t]
        \caption{Central mass and \textit{$\chi^{2}_\mathrm{red}$} values for the Keplerian and infalling fits employed in Figs.~\ref{fig:PIV} and \ref{fig:PIV_HDR}.}
        \label{table:PIV}
        \centering      
        \begin{tabular}{l l c c}
                \hline\hline
                Species & Fit & Central mass & \textit{$\chi^{2}_\mathrm{red}$} \\
                & & [M$_{\odot}$] & \\
                \hline
                \multicolumn{4}{c}{Disk-like structure} \\
                \hline
                C$^{17}$O & Keplerian & 2.2 $\pm$ 0.2 & 3.3 \\
                 & Infalling & 1.5 $\pm$ 0.2 & 3.3 \\
                H$^{13}$CO$^{+}$ & Keplerian & 2.2 $\pm$ 0.2 & 6.5 \\
                 & Infalling & 1.5 $\pm$ 0.2 & 15.1 \\           
                \hline
                \multicolumn{4}{c}{High-density region} \\
                \hline
                c$-$C$_{3}$H$_{2}$ & Keplerian &  2.2 $\pm$ 0.2 & 4.6 \\
                 & Infalling & 3.6 $\pm$ 0.3 & 2.4 \\
                 \hline
        \end{tabular}
\end{table}

\subsubsection{High-density region} 

\begin{figure}[t]
   \centering
      \includegraphics[width=.45\textwidth]{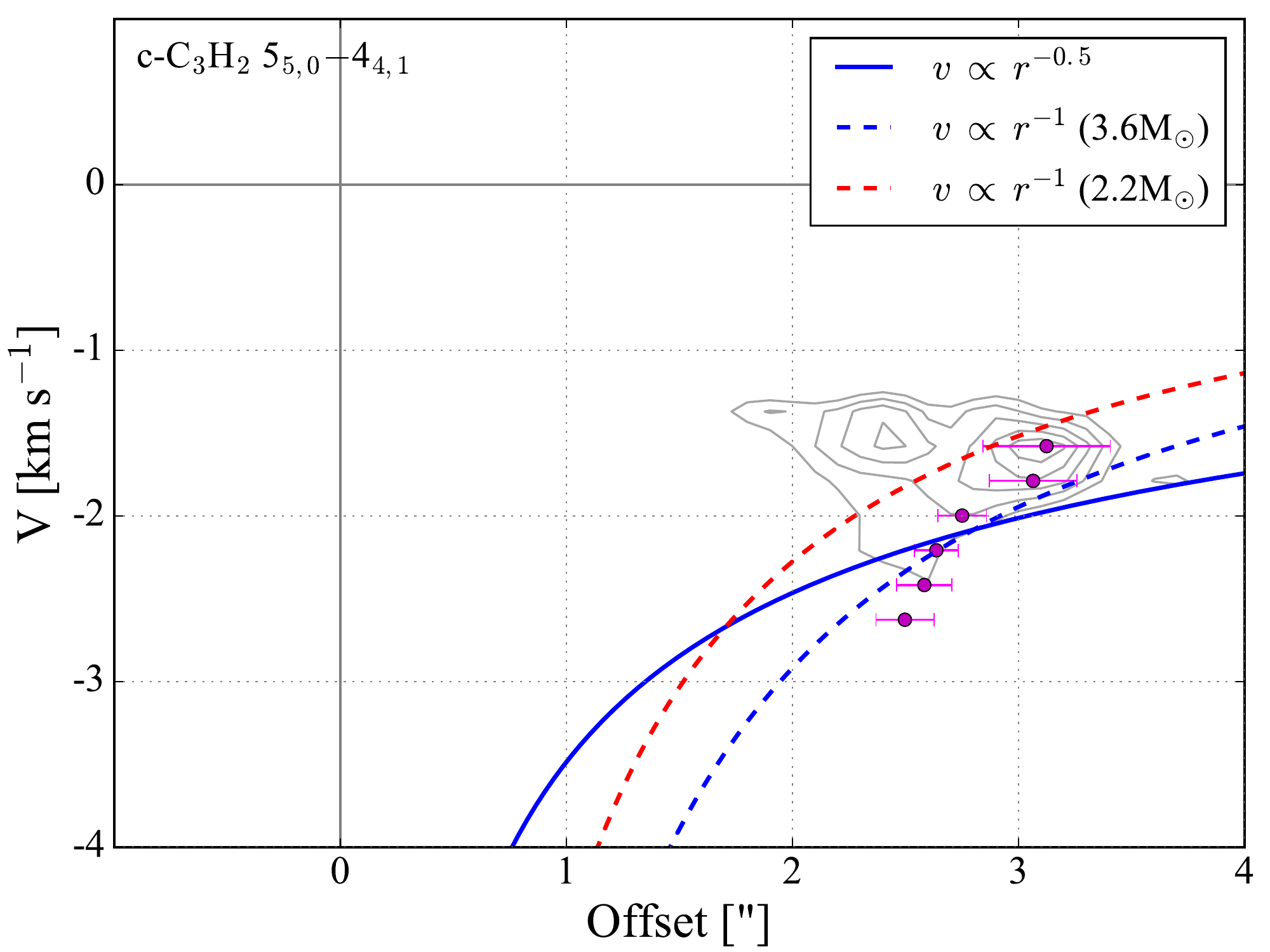}
      \caption[]{\label{fig:PIV_HDR}
          Position-velocity diagram for c$-$C$_{3}$H$_{2}$~5$_{5,0}$$-$4$_{4,1}$ towards the disk-like structure direction (PA = 234$\degr$) and centred at the position of the geometric centre. The magenta dots represent high-velocity (> 1.7~km~s$^{-1}$) emission peaks. Blue solid and dashed lines show the best fit for a Keplerian and an infalling velocity profile, respectively, while the red dashed line represents an infalling profile for a central mass of 2.2~M$_{\odot}$. The cut taken from the image data is shown in grey contours, ranging from 3$\sigma$ (1$\sigma$ = 10~mJy~beam$^{-1}$) to the maximum value. Each adjacent contour represents an increment of the 20$\%$ of the maximum value.
       }
\end{figure}

The PV diagram for c$-$C$_{3}$H$_{2}$ is shown in Fig.~\ref{fig:PIV_HDR}. In this case, the goodness of the infalling fit is better than the Keplerian one, with \textit{$\chi^{2}_\mathrm{red}$} values of 2.4 and 4.6, respectively (Table~\ref{table:PIV}). However, the central mass associated with the infalling fit is 3.6 $\pm$ 0.3~M$_{\odot}$, which is not consistent with the central mass of 2.2 $\pm$ 0.2~M$_{\odot}$ found from the Keplerian fit of C$^{17}$O and H$^{13}$CO$^{+}$. This implies that the c$-$C$_{3}$H$_{2}$ gas may not be following an infalling motion around the geometric centre and another velocity profile has to be considered for this molecular transition. 

The high-density region may also be a separate object with a different centre; however, if this were the case, we would expect compact continuum emission toward this region, contrary to what we see. Thus, it is more likely that the high-density region is related to the inner envelope or ambient cloud, and not to a third object.

\subsection{Disk mass}

The dust and gas masses are calculated from the continuum emission (Fig.~\ref{fig:continuum}) and from the integrated intensity of C$^{17}$O \textit{J}=3$-$2 (Fig.~\ref{fig:moments}a), respectively. For both cases, the calculated mass represents the entire system, i.e. both sources and the circumbinary disk contribution.

If the emission is optically thin, the total mass is

\begin{equation} 
    M = \frac{S_{\nu}d^{2}}{\kappa_{\nu}B_{\nu}(T)} \ ,
    \label{eq:Eq2}
\end{equation}

\noindent where \textit{S$_{\nu}$} is the surface brightness, \textit{d} the distance to the source, $\kappa$$_{\nu}$ the dust opacity and \textit{B$_{\nu}$(T)} the Planck function for a single temperature. 
For typical parameters of the dust temperature (30~K) and opacity at 0.87~mm \citep[0.0175~cm$^{2}$ per gram of gas; ][]{Ossenkopf1994}, commonly used for dust in protostellar envelopes and young disks in the millimetre regime \citep[e.g. ][]{Shirley2011}, the mass can be estimated as:

\begin{equation} 
      M_{0.87\mathrm{mm}} = 0.18 \ M_{\odot} \left( \frac{F_{0.87\mathrm{mm}}}{1 \ \mathrm{Jy}} \right) \left( \frac{d}{200 \ \mathrm{pc}} \right)^{2} \Bigg \lbrace \mathrm{exp} \left[ 0.55 \left( \frac{30 \ \mathrm{K}}{T} \right) \right] - 1 \Bigg \rbrace  \ ,
    \label{eq:Eq3}
\end{equation}

\noindent where \textit{M$_\mathrm{0.87mm}$} is the total (dust + gas) mass at 0.87~mm and \textit{F$_\mathrm{0.87mm}$} is the flux at 0.87~mm.

For the gas mass calculation, the intensity of C$^{17}$O \textit{J}=3$-$2 is integrated over a velocity range of 20~km~s$^{-1}$, obtaining a value of 5.31 $\pm$ 0.01~Jy~km~s$^{-1}$ for \textit{S$_{\nu}$} times the velocity interval (\textit{d$\varv$}), for emission above 5$\sigma$. Assuming a Boltzmann distribution with temperature \textit{T}, the total column density of C$^{17}$O, \textit{N$_\mathrm{total}$}(C$^{17}$O), is

\begin{equation} 
      N_\mathrm{{total}}(\mathrm{C^{17}O}) = \frac{4 \pi S_{v} d\varv}{\Omega A_{32} h c} \ Q(T) \ \mathrm{exp} \left( \frac{E_{3}}{k_\mathrm{B} T} \right), 
    \label{eq:Eq4}
\end{equation}

\noindent where $\Omega$ is the beam solid angle, \textit{A$_\mathrm{32}$} the Einstein coefficient for the transition \textit{J}=3$-$2, \textit{h} the Plank constant, \textit{c} the light speed, \textit{Q(T)} the partition function and \textit{E$_\mathrm{3}$}/\textit{k$_\mathrm{B}$} the upper level energy. Finally, the gas mass (\textit{M$_\mathrm{gas}$}) is

\begin{equation} 
      M_\mathrm{{gas}} = N_\mathrm{{total}}(\mathrm{C^{17}O}) \left[ \frac{\mathrm{CO}}{\mathrm{C^{17}O}} \right] \left[ \frac{\mathrm{H_{2}}}{\mathrm{CO}} \right] 2.8 m_\mathrm{H} \mathcal{A}, 
    \label{eq:Eq5}
\end{equation}
 
 \noindent where [CO/C$^{17}$O] and [H$_{2}$/CO] are relative abundances, 2.8~\textit{m$_\mathrm{H}$} is the mean molecular weight \citep{Kauffmann2008}, \textit{m$_\mathrm{H}$} the hydrogen mass and \textit{$\mathcal{A}$} the beam area. By assuming a C$^{17}$O abundance with respect to CO of 4.5 $\times$ 10$^{-4}$ \citep{Penzias1981} and a CO abundance with respect to H$_\mathrm{2}$ of 2.7 $\times$ 10$^{-4}$ \citep{Lacy1994}, gas masses are found for different temperatures. 
 
\begin{figure}[t]
   \centering
      \includegraphics[width=.49\textwidth]{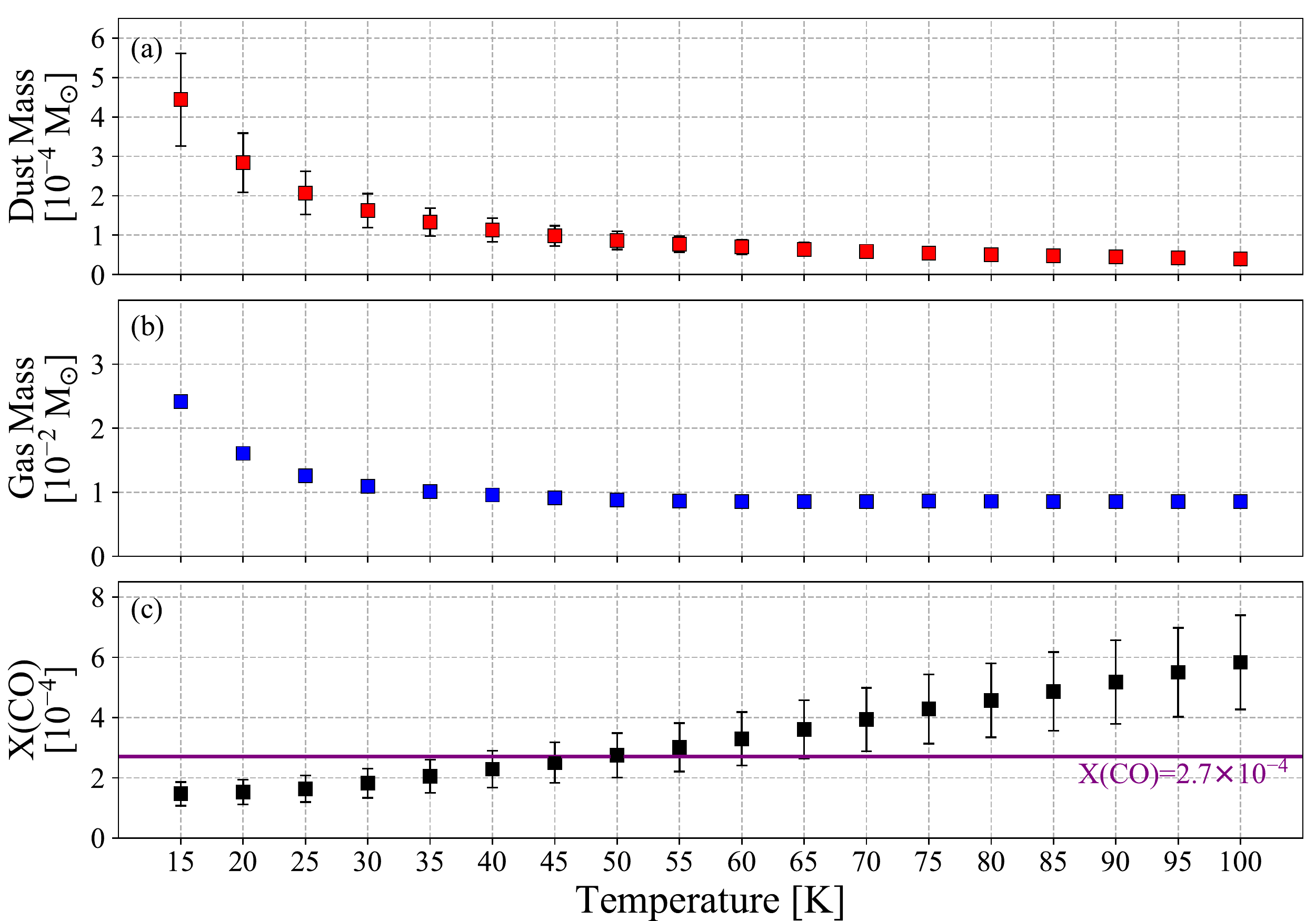}
      \caption[]{\label{fig:masses}
      Dust mass, gas mass, and values of CO abundances, for different temperatures and taken for the entire system (both sources and the circumbinary disk). The CO abundances are calculated by assuming a gas-to-dust ratio of 100. For gas masses (blue squares), the uncertainties are $\sim$0.3~$\%$ of the value and are imperceptible on the plot. The purple horizontal line shows the value of 2.7~$\times$~10$^{-4}$ for the abundance of CO with respect to H$_\mathrm{2}$.
      }
\end{figure}

Figure~\ref{fig:masses} shows the dust mass, gas mass, and CO abundances with respect to H$_\mathrm{2}$ for different temperatures. The dust and gas masses are calculated using Eqs.~\ref{eq:Eq3} and \ref{eq:Eq5}, respectively, while the CO abundances are calculated by isolating [CO/H$_{2}$] from Eq.~\ref{eq:Eq5} and assuming a gas-to-dust ratio of 100. The CO abundances are specifically inferred to test if a significant fraction of this molecule is frozen out in the cold midplane of the circumbinary disk, as would normally be expected. For comparison, the purple line shows the typical value of 2.7 $\times$ 10$^{-4}$ for the abundance of CO with respect to H$_\mathrm{2}$. Dust and gas masses increase as the temperature decreases, following an exponential behaviour, while a raise in the CO abundance is observed for higher temperatures. For a typical dust temperature of 30~K, a value of 1.1 $\times$ 10$^{-2}$~\textit{M$_{\odot}$} is obtained for the total mass and, by taking a gas-to-dust ratio of 100, a dust mass of 1.1 $\times$ 10$^{-4}$~\textit{M$_{\odot}$} is derived. We note that for all of the temperatures used in the calculations, CO abundances are comparable to the canonical value of 2.7 $\times$ 10$^{-4}$.

\subsection{Fits and abundances}

Towards the position of source B and source A, the central components from C$^{17}$O and H$^{13}$CO$^{+}$ show a slightly blue-shifted emission in the case of source B and a red-shifted emission for source A (Fig.~\ref{fig:spectra}a, b, d and e). This shift between both sources can be related to their proper motion, relative to the geometric centre. In order to obtain an accurate value of this shift, the central component of the spectra of both species is fitted with a Gaussian and the values are listed in Table~\ref{table:Gfit}. By comparing the velocity centroid for the same species, a difference of $\sim$1~km~s$^{-1}$ is seen between the sources. Apart form C$^{17}$O and H$^{13}$CO$^{+}$, SO$_{2}$ also shows emission towards source B and its emission is compact. The spectrum of SO$_{2}$ does not show a central component but presents blue- and red-shifted emission, thus, a two component Gaussian fit was employed. The blue component is more intense than the red one (20 vs. 9~mJy~beam$^{-1}$), however, the red-shifted emission presents a broader component (12 vs. 5~km~s$^{-1}$).

\begin{table*}[t]
        \caption{Gaussian fit values over the position of Oph-IRS67 A and B.}
        \label{table:Gfit}
        \centering
        \begin{tabular}{l l l c c c}
                \hline\hline
                Transition & Source & Component & Intensity & Centre & \textit{FWHM} \\
                & & & [mJy beam$^{-1}$] & [km s$^{-1}$] & [km s$^{-1}$]\\
                \hline
                C$^{17}$O \textit{J}=3$-$2 & Oph-IRS67B & Central & 63 $\pm$ 5 & -0.6 $\pm$ 0.1 & 2.4 $\pm$ 0.3 \\
                & Oph-IRS67A & Central & 90 $\pm$ 20 & 0.3 $\pm$ 0.2 & 1.4 $\pm$ 0.4 \\
                H$^{13}$CO$^{+}$ \textit{J}=4$-$3 & Oph-IRS67B & Central & 120 $\pm$ 10 & -0.9 $\pm$ 0.1 & 1.3 $\pm$ 0.1 \\
                & Oph-IRS67A & Central & 160 $\pm$ 20 & 0.2 $\pm$ 0.1 & 1.1 $\pm$ 0.2 \\
                SO$_{2}$ 18$_{4,14}$$-$18$_{3,15}$ & Oph-IRS67B & Blue & 20 $\pm$ 4 & -5.8 $\pm$ 0.4 & 5 $\pm$ 1 \\
                 & & Red & 9 $\pm$ 2 & 4 $\pm$ 2 & 12 $\pm$ 5 \\
                CH$_{3}$OH 7$_{-1,7}$$-$6$_{-1,6}$ E $^{(a)}$& --- & --- & < 11 $^{b}$ & --- & 2.4 $^{c}$ \\
                \hline
        \end{tabular}
        \tablefoot{Reported uncertainties are all 1$\sigma$. $^{(a)}$ Expected brightest transition at 100~K. $^{(b)}$ Value of 1$\sigma$. $^{(c)}$ Same value as C$^{17}$O.}
\end{table*}

The statistical equilibrium radiative transfer code \citep[RADEX; ][]{vanderTak2007} is used to estimate the optical depth ($\tau$) and column densities of C$^{17}$O, H$^{13}$CO$^{+}$, and SO$_{2}$, towards the position of source B. An upper limit for the CH$_{3}$OH column density is also determined by taking an intensity of 10~mJy~beam$^{-1}$ (1$\sigma$) and a \textit{FWHM} of 2.4~km~$^{-1}$ (same value as the \textit{FWHM} of the central component of C$^{17}$O; Table~\ref{table:Gfit}). The calculated values for $\tau$ and column density are listed in Table~\ref{table:RADEX}, and are obtained by assuming a H$_{2}$ number density of 1 $\times$ 10$^{7}$~cm$^{-3}$ (collision partner), two kinetic temperatures (\textit{T$_\mathrm{kin}$}) of 30 and 100~K, and a broadening parameter (\textit{b}) corresponding to the \textit{FWHM} value of each transition (see Table~\ref{table:Gfit}). The chosen value for the H$_{2}$ number density is related to the average value for the critical densities of the observed transitions (see Table~\ref{table:observations}). If we change the H$_{2}$ density by a factor of 10, the C$^{17}$O column density remains constant since its level populations are thermalized (the critical density for this transition is 3.5~$\times$~10$^{4}$~cm$^{-3}$; see Table~\ref{table:observations}). In the case of H$^{13}$CO$^{+}$, its abundance remains constant for an increasing H$_{2}$ density but may increase by a factor of 3 for a decreasing density (depending on the kinetic temperature). On the other hand, the SO$_{2}$ column density varies from 5~$\times$~10$^{15}$ to 1~$\times$~10$^{18}$ cm$^{-2}$ for an increasing density. Thus, the abundances presented in Table~\ref{table:RADEX} are lower limits for H$^{13}$CO$^{+}$ and approximate values for SO$_{2}$. The kinetic temperature of 30~K represents the lower limit for a rich molecular layer of the disk, warm enough that most molecules are in the gas phase, and that the chemistry is dominated by ion-neutral reactions \citep{Henning2013}. In addition, the kinetic temperature of 100~K represents the inner regions around the source, where a hot corino chemistry is expected. The optical thickness of C$^{17}$O and H$^{13}$CO$^{+}$ are close to 1 for a gas temperature of 30~K. For warmer gas (100~K), C$^{17}$O, H$^{13}$CO$^{+}$ , and CH$_{3}$OH emission are optically thin ($\tau < 1$), while SO$_{2}$ shows optically thick emission ($\tau > 1$). In the case of SO$_{2}$, a value of $\sim$16~K is obtained for the calculated excitation temperature (\textit{T$_\mathrm{ex}$}) of the model, suggesting that the line is sub-thermally excited.

The calculated relative abundances of HCO$^{+}$, SO$_{2}$ , and CH$_{3}$OH with respect to CO are shown in Table~\ref{table:abundances}, employing standard isotope ratios \citep{Wilson1999}. The obtained values are compared with abundance ratios observed for the prototypical Class 0 source, IRAS 16293-2422, taken from \citet{Schoier2002}. IRAS 16293-2422 is deeply embedded \citep[\textit{M$_\mathrm{env}$} $\sim$~4~M$_{\odot}$; ][]{Jacobsen2017} and more luminous \citep[$\textit{L}$~=~21~$\pm$~5~L$_{\odot}$; ][]{Jorgensen2016} than IRS 67, but it was chosen for the comparison since molecules such as CH$_{3}$OH and SO$_{2}$ were detected towards the system. \citet{Schoier2002} calculated molecular abundances for two different scenarios: \textit{(i)} using a constant molecular abundance relative to H$_{2}$, and \textit{(ii)} applying a jump in the fractional abundance at a temperature of 90~K. For scenario \textit{(ii)}, they present the abundances for the inner, dense, hot part of the envelope (\textit{T} > 90~K), and for the cooler, less dense, outer part of the envelope (\textit{T} < 90~K). Comparing our relative abundances with those from \citet{Schoier2002}, the [HCO$^{+}$]/[CO] abundance ratio agrees with a model with constant molecular abundance, the [SO$_{2}$]/[CO] abundance ratio is consistent with a relative abundance in the hot part of the envelope for a jump model, and the upper limit for the [CH$_{3}$OH]/[CO] abundance ratio is, at least, two orders of magnitude smaller than the obtained values for each model.

\begin{table*}[t]
        \caption{Retrieved parameters from RADEX by assuming a kinetic temperature (\textit{T$_\mathrm{kin}$}) and a broadening parameter (\textit{b}), for Oph-IRS67 B.}
        \label{table:RADEX}
        \centering
        \begin{tabular}{l l r c c r}
                \hline\hline
                Transition & Component & \textit{T$_\mathrm{kin}$} & \textit{b} & $\tau$ & \textit{N}(\textit{X})\\
                & & [K] & [km~s$^{-1}$] & & [cm$^{-2}$] \\
                \hline
                C$^{17}$O \textit{J}=3$-$2 & Central & 30 & 2.4 & 0.76 & 2.0 $\times$ 10$^{16}$ \\
                 &  & 100 & 2.4 & 0.14 & 2.3 $\times$ 10$^{16}$ \\
                H$^{13}$CO$^{+}$ \textit{J}=4$-$3 & Central & 30 & 1.3 & 1.02 & $\geq$~1.3 $\times$ 10$^{13}$ \\         
                 &  & 100 & 1.3 & 0.14 & $\geq$~8.3 $\times$ 10$^{12}$ \\
                SO$_{2}$ 18$_{4,14}$$-$18$_{3,15}$ & Blue &  100 & 5 & 1.97 & $\sim$~2.0 $\times$ 10$^{17}$ \\
                & Red & 100 & 12 & 2.04 & $\sim$~4.9 $\times$ 10$^{17}$ \\
                CH$_{3}$OH 7$_{-1,7}$$-$6$_{-1,6}$ E & --- & 100 & 2.4 & 0.02 & <~1.6 $\times$ 10$^{14}$ \\
                \hline
        \end{tabular}
        \tablefoot{Reported values are for a fixed H$_{2}$ density of 1~$\times$~10$^{7}$ cm$^{-3}$.}
\end{table*}

\begin{table*}[t]
        \caption{Comparison between calculated relative abundances with respect to CO for Oph-IRS67 B and values taken from \citet{Schoier2002} for the Class 0 protostar IRAS 16293-2422.}
        \label{table:abundances}
        \centering      
        \begin{tabular}{l c c c c}
                \hline\hline
                 Species & This work & \multicolumn{3}{c}{\citet{Schoier2002}} \\
                 \cline{3-5}
                 & & Constant abundance $^{a}$ & Jump (\textit{T} >  90~K) $^{b}$ & Jump (\textit{T} < 90~K) $^{c}$ \\
                \hline
                HCO$^{+}$ $^{d}$ & [4.8 $\times$ 10$^{-6}$ $-$ 2.2 $\times$ 10$^{-5}$] $^{e}$& \textbf{3.9 $\times$ 10$^{-5}$} & --- & --- \\
                SO$_{2}$ & [4.3 $\times$ 10$^{-3}$ $-$ 1.1 $\times$ 10$^{-2}$] $^{f}$& 1.7 $\times$ 10$^{-5}$ & \textbf{2.8 $\times$ 10$^{-3}$} & 1.3 $\times$ 10$^{-5}$ \\
                CH$_{3}$OH & < 3.5 $\times$ 10$^{-6}$ & 4.8 $\times$ 10$^{-5}$ & 2.8 $\times$ 10$^{-3}$ & 9.8 $\times$ 10$^{-6}$ \\    
                \hline
        \end{tabular}
        \tablefoot{$^{(a)}$ Value using a constant molecular abundance relative to H$_{2}$. $^{(b)}$ Value using a jump in the fractional abundance and obtained for the hot part of the envelope (\textit{T} > 90~K). $^{(c)}$ Value using a jump in the fractional abundance and obtained for the cooler part of the envelope (\textit{T} < 90~K). Highlighted values are closer to our calculations. $^{(d)}$ Assuming standard isotope ratios for C$^{17}$O ($^{16}$O/$^{17}$O = 2005) and H$^{13}$CO$^{+}$ ($^{12}$C/$^{13}$C = 69) from \cite{Wilson1999}. $^{(e)}$ For a range of temperatures, between 30 and 100~K. $^{(f)}$ For blue and red components.}
\end{table*}

\section{Discussion}
\subsection{Structure of Oph-IRS67}

The continuum emission (Fig.~\ref{fig:continuum}) shows the binary system, separated by 0\farcs71 $\pm$ 0\farcs01 (107 $\pm$ 2~AU), previously detected by \cite{McClure2010} at infrared wavelengths. They estimated a separation of 0.6$''$ ($\sim$ 90~AU) between the two sources and found that Oph-IRS67A (L1689S-A) is brighter than Oph-IRS67B (L1689S-B), while the opposite situation is observed in the sub-millimetre regime. The contrast between the sources at different wavelengths could be due to, (i) different evolutionary stages, i.e. Oph-IRS67A is more evolved than Oph-IRS67B, (ii) different dust grain sizes, (iii) different orientation of possible circumstellar disks, where the one associated with source A may be more face-on and the one related to source B more edge-on, (iv) different temperatures if the sources present distinct masses or one of them displays ongoing accretion bursts or, (v) the two sources lie differently in the circumbinary disk, where source B is more obscured than source A.

Since close binaries are expected to be coeval systems, options (i) and (ii) are less likely. Option (iii) is supported by SO$_{2}$ and C$^{17}$O emission, where both species show compact morphology around source B at high velocities. Such behaviour would be expected for an edge-on circumstellar disk, however this emission can also be related to accretion shocks \citep{Podio2015}. On the other hand, source A is associated with C$^{17}$O emission at low velocities but no SO$_{2}$ emission is observed towards this source. This supports the scenario where a possible circumstellar disk associated with source A is more face-on, where the shift in velocities is not as perceptible as in the edge-on case. Although a perpendicular misalignment between the two sources is statistically unlikely \citep[$\sim$ 2$\%$; ][]{Murillo2016}, this scenario will be consistent with source A being brighter than source B at infrared wavelengths, where the warm dust in the line of sight is not obscured by the cold dust. Option (iv) is plausible if the sources have different masses, however if one of the sources presents higher temperatures due to accretion bursts, we would expect to detect gas-phase CH$_{3}$OH around it. The arrangement proposed in option (v) agrees with the scenario suggested for IRS 43 \citep{Brinch2016}, where one of the sources lies behind the circumbinary disk and, therefore, is more obscured by the material along the line of sight. In the case of IRS 43, the two continuum sources have the same mass but differ in intensity by a factor of 5$-$10. For Oph-IRS67 AB, the intensity of the two sources differs by a factor of 5, however there are no constraints on the individual masses. 

The velocity difference between the two sources seen in C$^{17}$O and H$^{13}$CO$^{+}$ (Fig.~\ref{fig:spectra} and Table~\ref{table:Gfit}) is more consistent with emission arising from different regions than radial motion of both sources. If C$^{17}$O and H$^{13}$CO$^{+}$ trace radial motions, a better correlation between the parameters of their Gaussian fits is expected. This does not exclude that both sources are rotating around the centre of mass: indeed, such rotation is expected, but proper motion measurements are necessary to support this interpretation \citep[see, e.g.][]{Brinch2016}. 

At first glance, the weaker emission associated with the disk-like structure could be thought to be associated with a circumbinary disk, an inner infalling envelope or an outflow. However, the latter is ruled out since the system is already associated with an outflow perpendicular to the disk-like structure \citep{Bontemps1996}. On the other hand, neither the Keplerian nor the infalling fit can be ruled out from the PV diagrams of C$^{17}$O and H$^{13}$CO$^{+}$. A combination of both fits is more consistent with the data, where the large-scale structure is dominated by infalling motion and gas from inner regions follows a Keplerian profile \citep[e.g.][]{Harsono2014}. The presence of low-velocity points that seem to follow a linear distribution is expected for a Keplerian disk with finite radius. In addition, the lack of infalling signatures towards the perpendicular direction at small scales, agrees with the existence of a circumbinary disk associated with Keplerian motion around a central mass of 2.2 $\pm$ 0.2~M$_{\odot}$. Assuming that the major axes of the deconvolved continuum size represents the circumbinary disk diameter, its value of 620~AU is three times larger than the typical size of circumstellar disks around Class I sources, which is about 200~AU \citep{Harsono2014}. However, other detected circumbinary disks around Class I sources, such as IRS 43 \citep{Brinch2016} and L1551 NE \citep{Takakuwa2017}, show comparable sizes to that of Oph-IRS67 (diameters of $\sim$~650 and $\sim$~600 AU for IRS 43 and L1551 NE, respectively).

Given the low mass of the ambient core estimated from larger-scale SCUBA maps \citep[0.08~M$_{\odot}$; ][]{Jorgensen2008}, the two protostars can only accrete a small amount of additional material from here on. Assuming that both sources have similar masses, each of them would have a final mass of about 1~M$_{\odot}$, which is consistent with the total luminosity of the system \citep[4.0~L$_{\odot}$; ][]{Evans2009}.

The high-density region is enhanced in c$-$C$_{3}$H$_{2}$ and lies beyond the extension of the circumbinary disk; it stands out in intermediate blue-shifted velocities and there is a lack of red-shifted counterpart. This region also shows bright C$_{2}$H emission, but since this molecule also traces material from the vicinity of the protostars and is associated with red-shifted emission, its chemistry is not exclusively associated with the high-density region. The mentioned region could be related to, (i) a centrifugal barrier \citep[a transition zone within the disk where the kinetic energy of the infalling gas is converted into rotational energy; ][]{Sakai2014}, (ii) a region from the outflow cavity that is being irradiated by the sources, (iii) spiral-arm features formed by gravitational torques from the binary system or, (iv) ambient gas cloud or envelope that is infalling into the disk.

Option (i) was proposed by \cite{Sakai2014}, where they observe a drastic change in the chemistry (between c$-$C$_{3}$H$_{2}$ and SO) and argue that c$-$C$_{3}$H$_{2}$ follows an infalling motion, while SO shows a Keplerian profile. Therefore, the centrifugal barrier will be located in the discontinuous region between both molecular emission. In our case, a spatial difference is seen between C$^{17}$O and c$-$C$_{3}$H$_{2}$, however there is no evidence of a discontinuous change in the motion of the gas. The infalling fit for c$-$C$_{3}$H$_{2}$ is associated with a central mass that is not consistent with the value obtained from the Keplerian fit of C$^{17}$O, thus a centrifugal barrier cannot explain the presence of the high-density region in Oph-IRS67. Option (ii) was proposed by \cite{Murillo2018} for the Class 0 source VLA 1623. They observed a correlation between c$-$C$_{3}$H$_{2}$ and C$_{2}$H in a region located $\sim$300~AU from the source, and interpret that this emission is arising from the outflow cavity, where material is being irradiated by the star. Oph-IRS67 is associated with a large-scale outflow \citep{Bontemps1996} and it has been shown that the outflow opening angle widens with time \citep{Arce2006}. If the high-density region is related to the outflow cavity, an opening angle of $\sim$~60$\degr$ is needed for the radiation of the two stars to reach the region where c$-$C$_{3}$H$_{2}$ and C$_{2}$H are abundant. However, there is no evidence for red-shifted emission associated with c$-$C$_{3}$H$_{2}$ and a symmetric blue-shifted counterpart on the other side of the postulated outflow cavity. Option (iii) was observed for the protobinary system L1551 NE \citep{Takakuwa2017}, where velocity gradients suggest the presence of expanding gas motions in the arms. This was observed in C$^{18}$O emission and the velocity gradient was found between blue- and red-shifted emission from transverse directions with respect to the circumbinary disk. For IRS 67, the search for a velocity gradient will be incomplete without the red-shifted counterpart, however, this possibility cannot be ruled out. Finally, option (iv) was predicted by circumstellar disk formation simulations \citep[e.g. ][]{Harsono2015a}: Shocks are generated when the infalling material from the ambient gas or envelope reaches the disk, and may induce the enhancement of some species, such as c$-$C$_{3}$H$_{2}$ and C$_{2}$H in the case of IRS 67. The presence of a red-shifted counterpart is expected for this scenario, however, the red-shifted emission may be obscured by the circumbinary disk material. Further modeling efforts and observations of multiple species would be required to clarify the nature of the high-density region.

C$^{17}$O and SO$_{2}$ show compact and high-velocity emission only around source B. This emission arises from a region with a diameter of $\sim$60~AU and could be related to a circumstellar disk or to accretion shocks. If circumstellar disks are present around the sources, they are unresolved in our data and observations with better resolution are necessary. Resolving possible circumstellar disks will be necessary to evaluate whether or not they are perpendicularly misaligned and to estimate individual masses.

\subsection{Methanol abundances}

The rest frequencies of methanol transitions \textit{J$_{k}$}=7$_{k}$$-$6$_{k}$ are covered by one of the spectral windows (see Table~\ref{table:observations}) but not detected. The lack of emission indicates an absence of methanol in the gas phase, or a low abundance with respect to H$_{2}$ ( < 10$^{-10}$$-$10$^{-9}$; see Table~\ref{table:abundances}). The stringent upper limit is similar to what was found towards R~CrA IRS7B in ALMA Cycle~0 observations by \cite{Lindberg2014}, where they found relatively low CH$_{3}$OH abundances ( < 10$^{-10}$$-$10$^{-8}$) compared to what is seen in the innermost \textit{T} > 90$-$100~K regions of some Class 0 sources, where the abundances are 10$^{-8}$$-$10$^{-7}$ \citep{Jorgensen2005a}. \cite{Lindberg2014} suggest either  that methanol formation is suppressed by the high temperature in the cloud, preventing significant CO to freeze-out, or the presence of an extended circumstellar disk. The extended disk would dominate the mass budget in the innermost regions ($\lesssim$~100~AU) of the protostellar envelope and generate a flattened density profile, that is, \textit{n}~$\propto$~\textit{r$^{-1.5}$} for \textit{r}~$\gtrsim$~100~AU and \textit{n}~=~constant for \textit{r}~$\lesssim$~100~AU \citep{Lindberg2014}. For a flat profile, a lower column density of material with high temperatures is expected. Since CH$_{3}$OH desorbs from dust grains at a temperature of $\sim$100~K, less material at high temperatures would imply a lower column density for gas-phase CH$_{3}$OH. Therefore, the flattened profile associated with the disk evolution would explain the low methanol column density seen toward circumstellar disks around Class I sources \citep[e.g.][]{Lindberg2014}. These observations thus point to a scenario in which gas-phase methanol column densities are low in the Class I stage, as opposed to the younger Class 0 stage where methanol is often present and detected \citep[e.g.][]{Jorgensen2005b, Maret2005, Kristensen2010}.

\subsection{Disk mass and temperature structure}

Independent from temperature, the CO abundance is found to be comparable with the canonical value of 2.7 $\times$ 10$^{-4}$. This implies that freeze-out of C$^{17}$O onto dust grains in the disk midplane or conversion of CO into other species \citep[e.g. ][]{Anderl2016} is not significant, assuming thin dust and C$^{17}$O emission. Figure~\ref{fig:continuum} shows that the continuum emission arises mainly from source B and presents a minor contribution from source A and the circumbinary disk. On the other hand, the gas emission from C$^{17}$O (Fig.~\ref{fig:moments}a) stands out toward source B and has an important contribution from the circumbinary disk, adding more surface area with significant contribution to the total gas mass. A similar case was seen for the proto-binary system IRS 43 \citep{Brinch2013}, where HCO$^{+}$ shows strong lines and extended emission while the continuum is relatively weak and compact. They also studied the case of IRS 63 that shows a brighter continuum and weak HCO$^{+}$ emission, the opposite situation to IRS 43. The disk associated to IRS 63 is more massive than that of IRS 43 \citep[0.053 vs. 0.0081~M$_{\odot}$; ][]{Jorgensen2009} and could explain the difference seen in HCO$^{+}$. The more massive the disk, the stronger the shielding would be, and thus, the colder the midplane where molecules can effectively freeze-out. Oph-IRS67 and IRS 43 may present the same situation, where the disk mass is not enough to provide an efficient shielding and C$^{17}$O (for the case of Oph-IRS67) is hardly affected by freeze-out. This means that most of the C$^{17}$O is in the gas phase and a significant fraction of the disk has temperatures $\geq$ 30~K (temperature at which CO sublimates from dust grains). In addition to the radiation field of both sources, the presence of an envelope can be regarded as an extra heating source, where the envelope serves as a blanket and the circumbinary disk remains warm \citep[e.g. ][]{Harsono2015b}.  

From C$^{17}$O emission and lack of methanol emission, a range of temperatures between 30 and 100~K is suggested for the circumbinary disk gas content. This is a wide range, however other molecular transitions with sublimation temperatures between 30 and 100~K are needed in order to constrain the temperature profile. In particular, the detection of formaldehyde (H$_{2}$CO) towards IRS 67 \citep{Lindberg2017} agrees with this range of temperatures, since H$_{2}$CO can be formed in the gas-phase or via grain surface chemistry. In the gas phase, CO reacts with H$_{2}$ to form H$_{2}$CO, and temperatures above 30~K are needed in order to prevent CO freeze-out, while a temperature of $\sim$50~K is required for H$_{2}$CO to sublimate from the grain mantles.

\section{Summary}

This work presents high-angular-resolution (0\farcs4, $\sim$60~AU) ALMA observations of the proto-binary system Oph-IRS67AB and its associated circumbinary disk. The continuum emission is analysed, together with molecular species such as C$^{17}$O, C$^{34}$S, H$^{13}$CO$^{+}$, SO$_{2}$, C$_{2}$H, c$-$C$_{3}$H$_{2}$ and CH$_{3}$OH. The main results of this paper are provided below. 

The flattened structure seen in the continuum proves the existence of a circumbinary disk around sources A and B, for which a diameter of $\sim$620~AU is estimated, assuming an inclination of $\sim$80$\degr$. The kinematic analysis supports a scenario where a Keplerian profile is associated with the gas motion of the circumbinary disk around a 2.2~\textit{M$_{\odot}$} central mass, however other velocity profiles cannot be ruled out from this analysis.

The contrast between sources A and B at different wavelengths is more likely to be a consequence of different temperatures between both sources or the position of the two sources within the circumbinary disk.

The high-density region shows strong emission from carbon chain molecules (C$_{2}$H and c$-$C$_{3}$H$_{2}$) and may be related to the inner envelope material (i.e. outflow cavity, spiral-arm structures or infalling material from the envelope to the circumbinary disk).

The SO$_{2}$ transition is optically thick ($\tau$ $\approx$ 2), shows compact emission, and is detected only around source B. The emission arises from a region of $\sim$60~AU in diameter, which is comparable with the beam size ($\sim$45~AU $\times$ 60~AU). The width of the line ($\sim$20~km~s$^{-1}$) may be related with the presence of an edge-on circumstellar disk or with accretion shocks. The absence of SO$_{2}$ emission towards source A is an intriguing fact and observations with better resolution are needed.

The lack of methanol emission suggests an upper limit between 10$^{-10}$ and 10$^{-9}$ for the abundance of CH$_{3}$OH. This can be explained by the presence of the disk, where the bulk of the material is dominated by the flattened structure in the inner regions and less material is subjected to high temperatures ($\sim$100~K), preventing CH$_{3}$OH sublimation from dust grains.

For different temperatures, calculated CO abundances are comparable to the canonical value of 2.7 $\times$ 10$^{-4}$, suggesting a non-effective shielding of the dust and, thus, no significant freeze-out of CO in the midplane. Considering that CO sublimates from dust grains at $\sim$30~K, and the lack of methanol detection, a temperature between 30 and 100~K characterises the bulk of the circumbinary disk.

The main conclusions of this work are the detection of the circumbinary disk and the clear difference seen by the chemistry between the high-density region and the circumbinary disk, where different physical processes are taking place in each of them. The high-angular-resolution observations make it possible to distinguish and differentiate the two regions. Future observation of other molecules and more transition from c$-$C$_{3}$H$_{2}$ will provide more information about the high-density region, while higher-angular-resolution observations are needed in order to resolve the cimcumstellar disks around each source and explain the chemical differences seen between them. Finally, Oph-IRS67AB is an ideal candidate for proper motion studies and planet formation, not only around circumstellar disks, but also around circumbinary disks.

\begin{acknowledgements}

We thank the anonymous referee for a number of good suggestions helping to improve the presentation of this work. This paper makes use of the following ALMA data: ADS/JAO.ALMA\#2013.1.00955.S. ALMA is a partnership of ESO (representing its member states), NSF (USA) and NINS (Japan), together with NRC (Canada), NSC and ASIAA (Taiwan), and KASI (Republic of Korea), in cooperation with the Republic of Chile. The Joint ALMA Observatory is operated by ESO, AUI/NRAO and NAOJ. The group of JKJ acknowledges support from the European Research Council (ERC) under the European Union's Horizon 2020 research and innovation programme (grant agreement No 646908) through ERC Consolidator Grant ``S4F''. Research at the Centre for Star and Planet Formation is funded by the Danish National Research Foundation. 

\end{acknowledgements}

\bibliographystyle{aa} 
\bibliography{References} 

\begin{appendix}
\section{Channel maps of individual molecular transitions}

\begin{figure*}[!t]
   \centering
      \includegraphics[width=.98\textwidth]{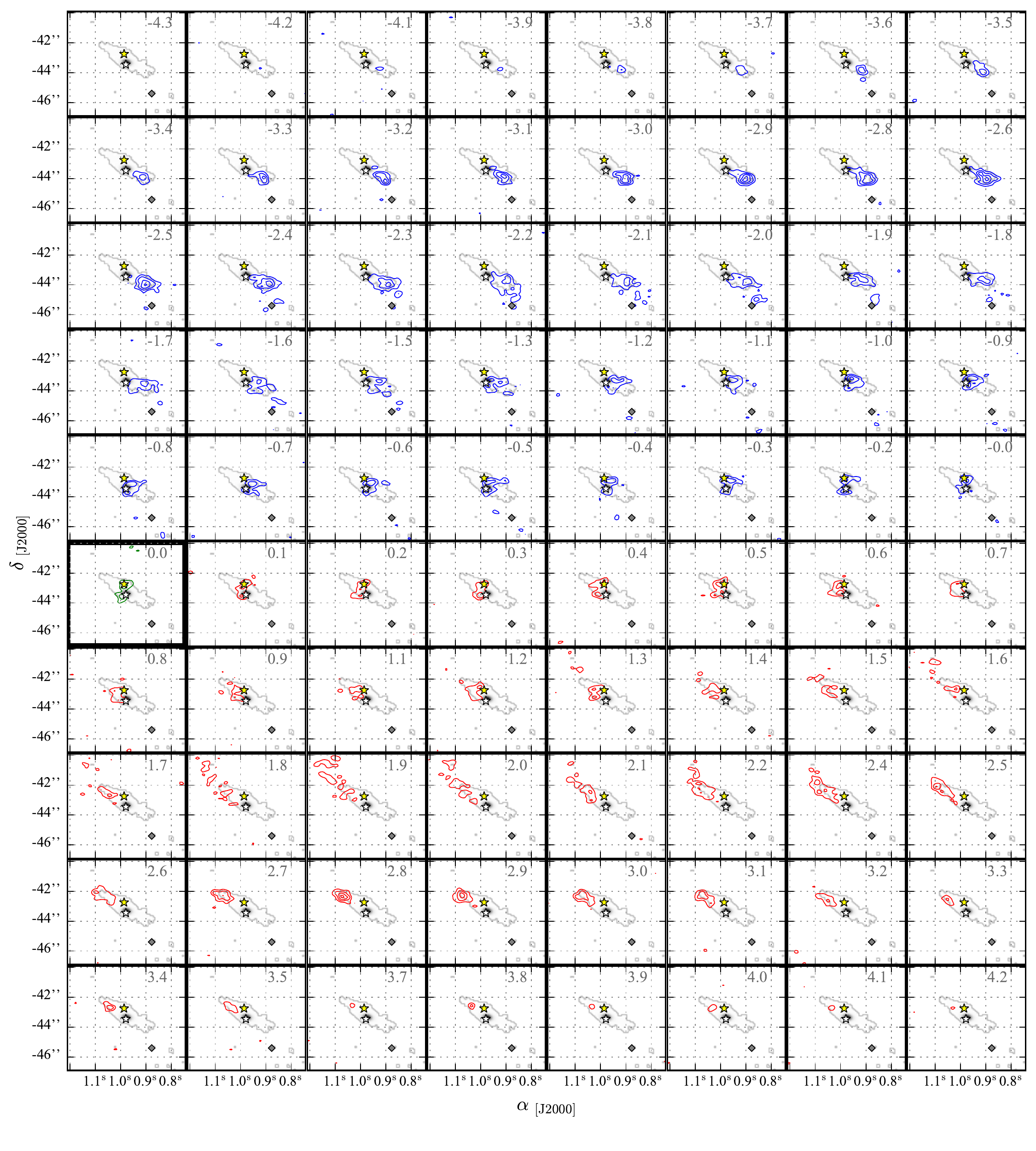}
      \caption[]{\label{fig:chans_C17O}
      Velocity channel maps for C$^{17}$O \textit{J}=3$-$2 (contours), superposed on the 0.87~mm continuum image (grey-scale; Fig.~\ref{fig:continuum}). The highlighted panel with green contours represents the systemic velocity of 4.2~km~s$^{-1}$. Blue and red contours indicate blue- and red-shifted emission with respect to the systemic velocity, respectively. In each panel, the contours start at 4$\sigma$ ($\sigma$ = 10~mJy~beam$^{-1}$ per channel), following a step of 3$\sigma$, and the mean velocity of each channel is also presented in units of km~s$^{-1}$. The yellow and white stars show the positions of Oph-IRS67A and B, respectively. The grey diamond denotes the location of the high-density region.
      }
\end{figure*}

\begin{figure*}[t]
   \centering
      \includegraphics[width=.98\textwidth]{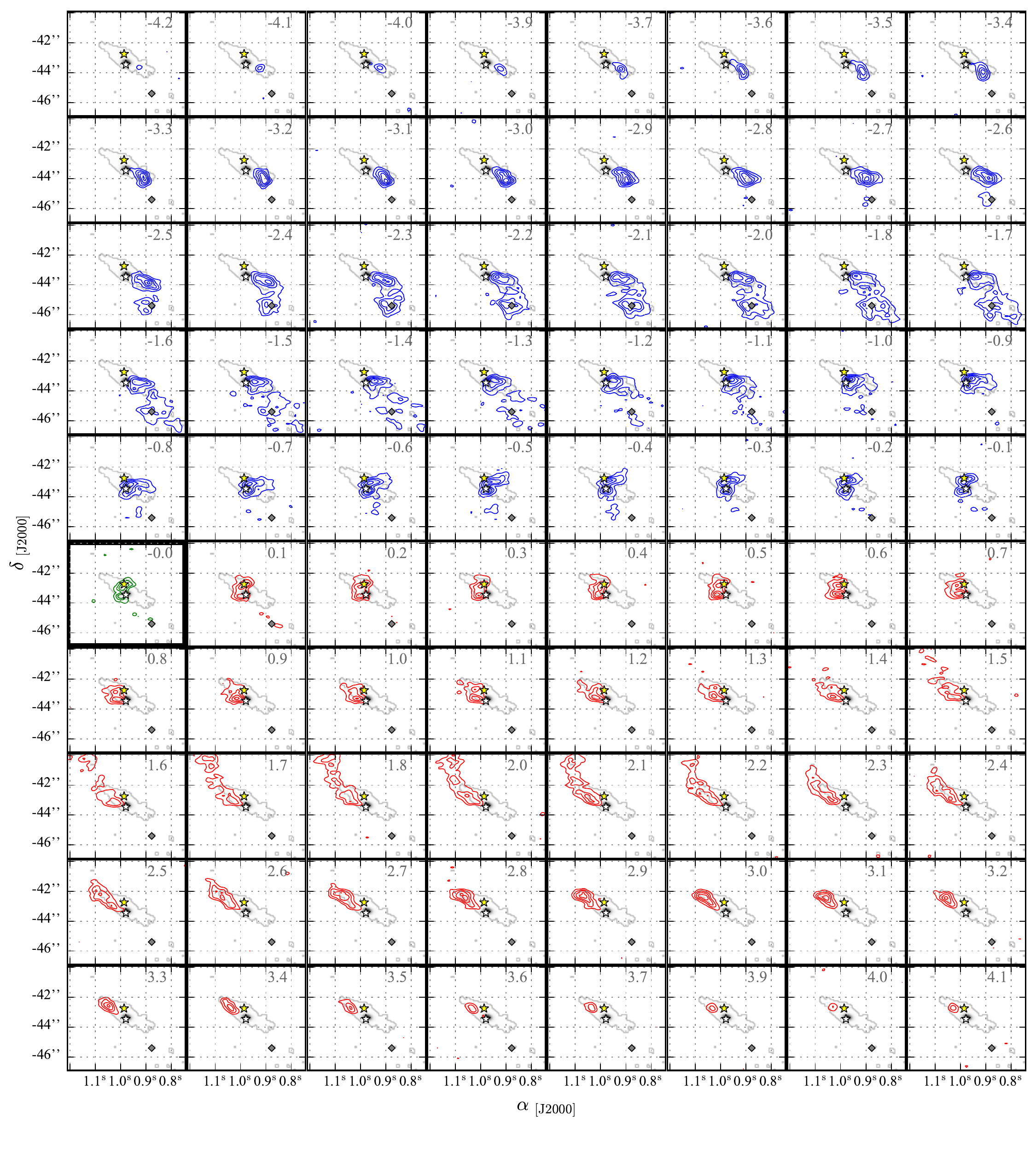}
      \caption[]{\label{fig:chans_H13CO+}
      As in Fig.~\ref{fig:chans_C17O} but for H$^{13}$CO$^{+}$ \textit{J}=4$-$3, where contours start at 4$\sigma$ and follow a step of 4$\sigma$.
      }
\end{figure*}

\begin{figure*}[t]
   \centering
      \includegraphics[width=.98\textwidth]{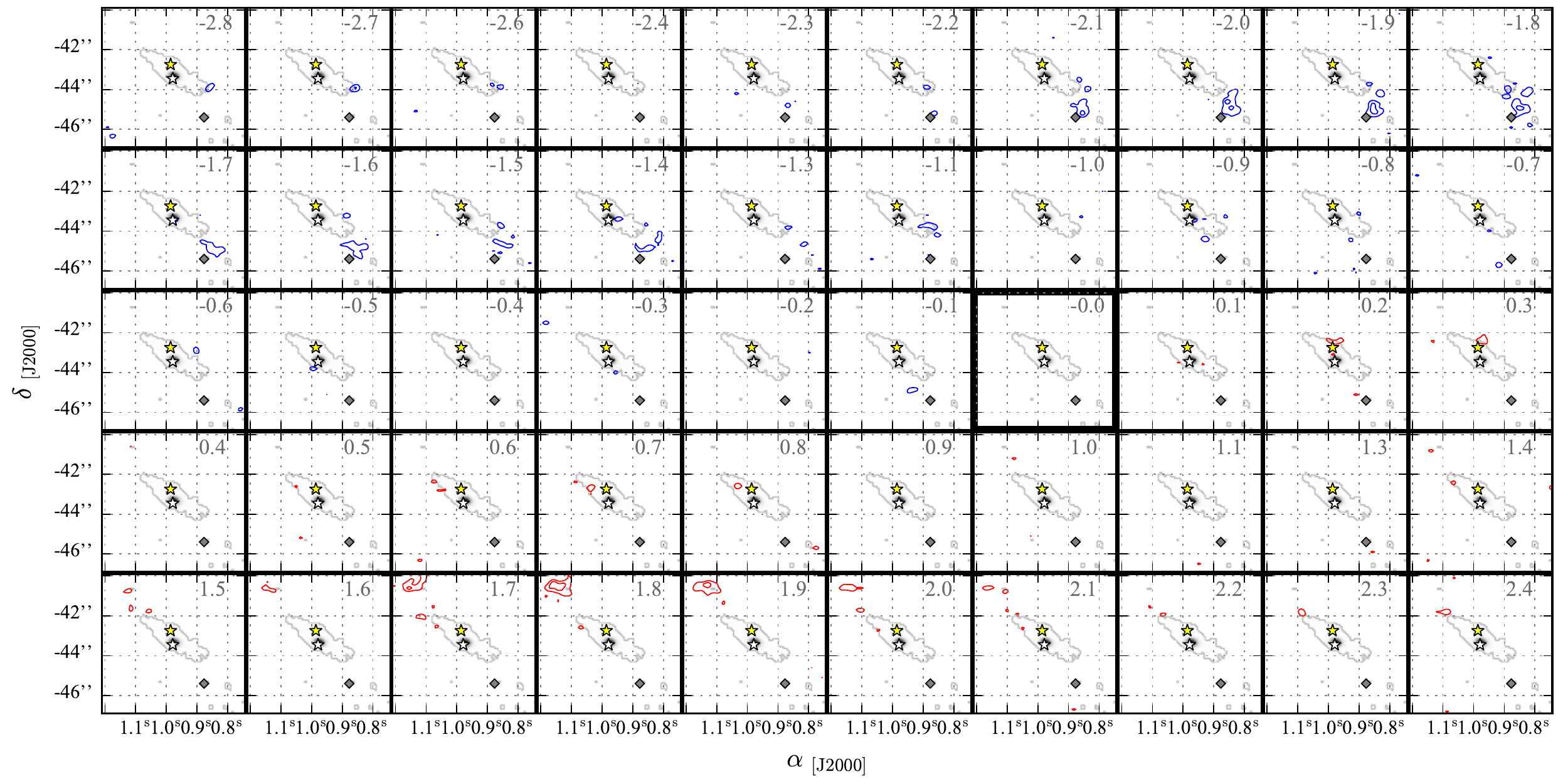}
      \includegraphics[width=.98\textwidth]{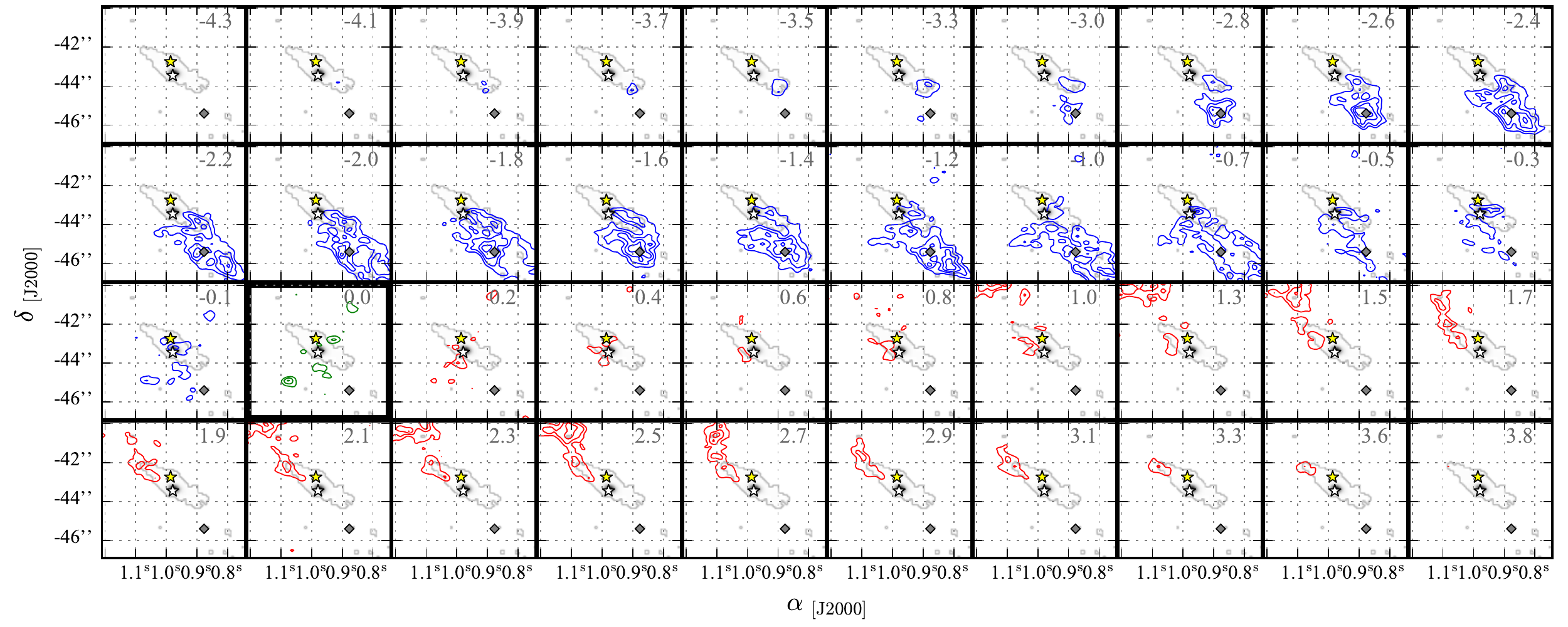}
      \includegraphics[width=.98\textwidth]{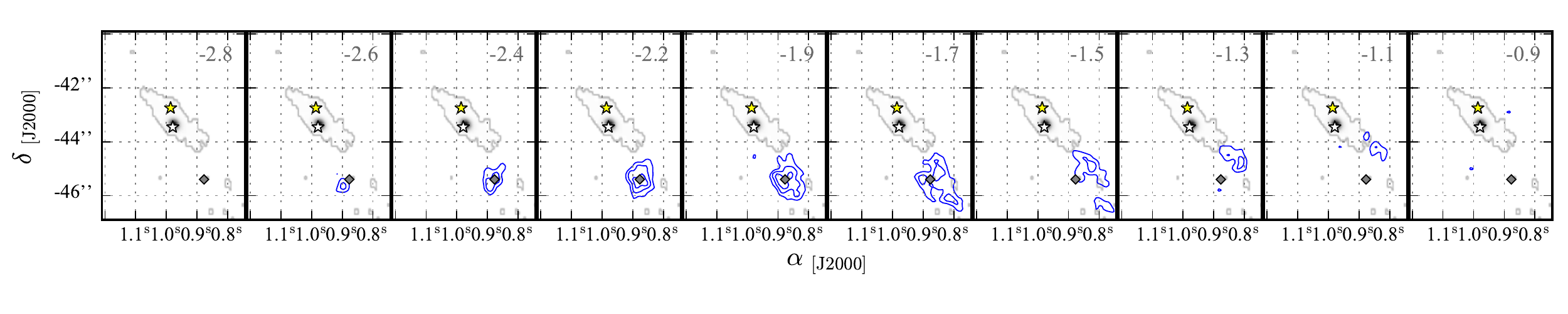}
      \caption[]{\label{fig:chans_rest}
      As in Fig.~\ref{fig:chans_C17O} but for C$^{34}$S \textit{J}=7$-$6 (\textit{upper frame}), the doublet C$_{2}$H \textit{N}=4$-$3, \textit{J}=9/2$-$7/2 (\textit{central frame}) and c$-$C$_{3}$H$_{2}$ 5$_{5,0}$$-$4$_{4,1}$ (\textit{lower frame}).
      }
\end{figure*}

For a more detailed inspection, channel maps for C$^{17}$O, H$^{13}$CO$^{+}$, C$^{34}$S, the doublet C$_{2}$H \textit{N}=4$-$3, \textit{J}=9/2$-$7/2 and c$-$C$_{3}$H$_{2}$ 5$_{5,0}$$-$4$_{4,1}$ are shown in Figs.~\ref{fig:chans_C17O},~\ref{fig:chans_H13CO+} and~\ref{fig:chans_rest}. The main differences between these channel maps and Figs.~\ref{fig:moments} and~\ref{fig:moments_bis} are, (i) the C$^{17}$O emission appears more extended than the continuum emission around $\pm$~2~km~s$^{-1}$, (ii) C$^{17}$O and H$^{13}$CO$^{+}$ have significant emission around sources A and B for low velocities (between -1 and 1~km~s$^{-1}$) and, (iii) there is no C$_{2}$H emission at the positions of either sources.

\end{appendix}

\end{document}